\documentclass[prb,10pt, reprint, groupedaddress, superscriptaddress]{revtex4-2}

\usepackage[usenames,dvipsnames]{xcolor}
\usepackage{amssymb}
\usepackage{graphicx}
\usepackage{amsmath}
\usepackage{txfonts}
\usepackage[bookmarks=true,colorlinks,citecolor=blue,urlcolor=blue]{hyperref}
\usepackage{braket}
\usepackage{babel}
\usepackage{blindtext}
\usepackage[compat=1.1.0]{tikz-feynman}
\usepackage[normalem]{ulem}
\usepackage{physics}
\usepackage{tikz}
\usepackage{mathtools}
\usepackage{pgfplots}

\begin{document}

\newcommand{\TUM}{\affiliation{Technical University of Munich, TUM School of Natural Sciences, Physics Department, 85748 Garching, Germany}}
\newcommand{\MCQST}{\affiliation{Munich Center for Quantum Science and Technology (MCQST), Schellingstr. 4, 80799 M{\"u}nchen, Germany}}

\def\papertitle{Finite Temperature Entanglement Negativity of Fermionic Symmetry Protected Topological Phases and Quantum Critical Points in One Dimension}
\title{\papertitle}

 \author{Wonjune Choi} \TUM \MCQST
 \author{Michael Knap} \TUM \MCQST
 \author{Frank Pollmann} \TUM \MCQST

\date{\today}

\begin{abstract}
We study the logarithmic entanglement negativity of symmetry-protected topological (SPT) phases and quantum critical points (QCPs) of one-dimensional noninteracting fermions at finite temperatures.
In particular, we consider a free fermion model that realizes not only quantum phase transitions between gapped phases but also an exotic topological phase transition between quantum critical states in the form of the fermionic Lifshitz transition.
The bipartite entanglement negativity between adjacent fermion blocks reveals the crossover boundary of the quantum critical fan near the QCP between two gapped phases.
Along the critical phase boundary between the gapped phases, the sudden decrease in the entanglement negativity signals the fermionic Lifshitz transition responsible for the change in the topological nature of the QCPs.
In addition, the tripartite entanglement negativity between spatially separated fermion blocks counts the number of topologically protected boundary modes for both SPT phases and topologically nontrivial QCPs at zero temperature.
However, the long-distance entanglement between the boundary modes vanishes at finite temperatures due to the instability of SPTs, the phases themselves.
\end{abstract}

\maketitle

\section{Introduction \label{sec: intro}}

As a platform to realize rich families of quantum phases of matter, a free fermion system demonstrates that simplicity does not necessarily imply triviality.
Even without interactions, fermions can have nontrivial ground states such as symmetry-protected topological (SPT) phases and quantum critical points (QCPs) \cite{Haldane, Kane_TI, Kane_TI_RMP, Qi_TIreview, Chiu_SPTclassify}.
Several strongly correlated systems, such as the transverse field Ising chain \cite{LSM_Ising}, are exactly solved by mapping the interacting problems to free fermion systems, yet they serve as paradigmatic models for symmetry breaking and SPT phases.
Hence, free fermion models are simple yet representative Hamiltonians to explore novel quantum phases and critical points.

An important characteristic of nontrivial quantum phases is the entanglement structure \cite{area_law_EE, Frank_1Dfermion, Chen_SPT, Zeng_SPT_entropy, Zeng_SSB_entropy}.
In particular, fermionic SPTs are differentiated from the trivial state due to their topological structures of short-range entanglement protected by symmetries \cite{Frank_1Dfermion, Chen_SPT, Zeng_SPT_entropy}.
Although the SPTs protected by an on-site symmetry are unstable at finite temperatures \cite{finiteT_SPT_unstable}, they are expected to leave signatures of the ground state entanglement at low enough temperatures \cite{Grover_singularity_negativity, Grover_entanglement_length, Grover_topological_negativity_finiteT}.
Detecting entanglement experimentally is a long-sought-after goal.
However, since experiments are typically done at finite temperatures, identifying the finite temperature signature of the quantum entanglement for the SPTs and the QCPs of noninteracting fermions is a pertinent problem to understand not only the free fermion systems themselves but also mean-field theories described by fermionic Gaussian states as trial wavefunctions.

This work investigates the mixed-state entanglement of fermionic topological phases and QCPs at finite temperatures.
Specifically, we focus on the three-parameter family of one-dimensional free fermion Hamiltonians \cite{Ruben_topology_fermion, Ruben_1dSPT}, which is the fermionic dual of the extended XZX cluster-Ising model \cite{cluster_Ising, oneway_QC} via the Jordan-Wigner transformation \cite{Jordan-Wigner, Ruben_1dSPT}:
\begin{align}
H &= \frac{i}{2} \sum_{n=1}^L  \Big( g_2 c_{2n-3}c_{2n}
- g_1 c_{2n+1}c_{2n}
+g_0 c_{2n-1}c_{2n}\Big),
\label{eq: free_fermion}
\end{align}
where $c_j$ are Majorana fermions satisfying $c_j c_k + c_k c_j = 2\delta_{jk}$.
Importantly, the Hamiltonian is solvable and allows us to explore not only QCPs between gapped SPTs \cite{Kitaev_pwave, SSH, Ruben_1dSPT} but also an unusual topological phase transition (the so-called quantum Lifshitz transition) between the QCPs \cite{Ruben_topology_fermion}.
Like gapped SPTs, there exist topologically distinct families of QCPs described by conformal field theories (CFTs); when symmetries are not explicitly or spontaneously broken, not every pair of the quantum critical states in the same universality class can be adiabatically connected without crossing a nonconformal multicritical point or opening up the bulk gap \cite{Ruben_topology_fermion, Ruben_SETcriticality}.
The chosen free fermion model permits the systematic study of these so-called gapless SPTs or symmetry-enriched criticalities \cite{Ruben_SETcriticality}, whose entanglement properties are less understood than those of gapped SPTs.

To quantify the many-body entanglement of the thermal mixed states, we calculate logarithmic entanglement negativity \cite{Vidal_negativity, negativity_monotone, Shapourian_partial_TR, Shapourian_fermion_negativity}.
Entanglement entropy (EE), the standard measure of the bipartite entanglement of a pure state, has limited utilities for a thermal state because it is sensitive to both thermal and quantum fluctuations.
However, the entanglement negativity is only sensitive to genuine quantum correlations and blind to thermal fluctuations \cite{Grover_singularity_negativity, Grover_entanglement_length, Grover_topological_negativity_finiteT}.
While the entanglement negativity for the spin or bosonic systems \cite{Vidal_negativity, Vidal_QCPentanglement} can be naturally defined from the positive partial transpose (PPT) criterion \cite{Peres_separability, Horodecki_separability}, the partial transpose of the fermionic density matrix requires careful consideration because of anticommuting exchange statistics \cite{Shapourian_fSPT_invariant, Shiozaki_fSPT_invariant, Eisler_partial_transpose, Shapourian_partial_TR, Shapourian_fermion_negativity}.
There have been two different proposals for the partial transpose of fermionic density operators and the corresponding \emph{fermionic} negativity \cite{Eisler_partial_transpose, Shapourian_partial_TR}, and we adopt the definition proposed in Ref.~\cite{Shapourian_partial_TR} throughout the paper.

The finite temperature entanglement negativity has been extensively studied for spontaneous symmetry broken phases \cite{Grover_entanglement_length, Grover_singularity_negativity}, topological orders \cite{Grover_topological_negativity_finiteT, finiteTnegativity_toric_code}, conformal field theories \cite{Calabrese_finiteTnegativity}, and free fermions \cite{Shapourian_finiteTnegativity}.
However, most of the previous works focused on the universal finite temperature scaling and singularities of the entanglement negativity at QCPs \cite{Grover_entanglement_length, Grover_singularity_negativity, Grover_topological_negativity_finiteT, finiteTnegativity_toric_code, Calabrese_finiteTnegativity} with the conventional definition of entanglement negativity for bosons and spins \cite{Vidal_negativity}.
Thus, further research with the refined definition of entanglement negativity for microscopic fermionic systems \cite{Eisler_partial_transpose, Shapourian_partial_TR, Shapourian_fermion_negativity} is useful to better understand the bipartite entanglement of fermionic quantum matter.

Here, we particularly focus on the fermionic negativity near the QCPs between the gapped phases and the fermionic Lifshitz tricritical point between the gapless states.
On the one hand, near the QCP between the gapped SPTs, the fermionic negativity exhibits a clear structure of quantum critical fan at finite temperatures.
The fermionic negativity of thermal states inside the quantum critical fan decays logarithmically $\sim \ln \left(1/T\right)$ at low temperatures.
On the other hand, there is a pronounced sudden decrease of the fermionic negativity at the fermionic Lifshitz tricritical point between the two topologically distinct Ising CFTs.
We analytically show that the fermionic Lifshitz tricritical point has constant ground state entanglement regardless of the system size and numerically confirm that the fermionic negativity remains approximately constant at low temperatures.
With the tripartite geometry of the two spatially separated intervals, we find that the fermionic negativity can distinguish distinct gapped and gapless topological ground states from the long-distance entanglement between the topologically protected localized boundary modes.

The remainder of the paper is organized as follows. 
In Sec.~\ref{sec: model}, we summarize the SPTs and QCPs of the free fermion model in Eq.~(\ref{eq: free_fermion}).
Sec.~\ref{sec: negativity} provides a self-contained review of the logarithmic entanglement negativity proposed in Ref.~\cite{Shapourian_partial_TR} and outlines the calculation scheme for a system of noninteracting fermions.
In Sec.~\ref{sec: adjacent}, we present our main results about the bipartite fermionic negativity of the SPTs and QCPs at finite temperatures.
Sec.~\ref{sec: disjoint} discusses the fermionic negativity between two spatially separated intervals and illustrates how the topologically protected edge modes contribute to the fermionic negativity.
We conclude our paper in Sec.~\ref{sec: conclusion} with a discussion and an outlook.

\section{Model \label{sec: model}}

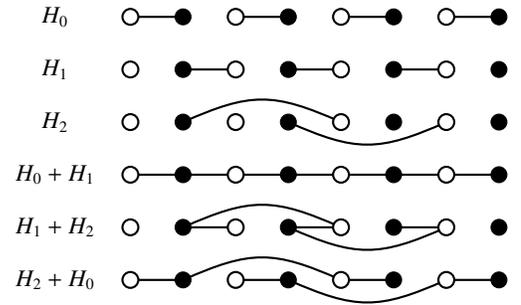
\begin{figure}[h]
\begin{tikzpicture}

\draw[] (-1, 3.5) node { $H_0$ };
\draw[] (-1, 2.8) node { $H_1$ };
\draw[] (-1, 2.1) node { $H_2$ };
\draw[] (-1, 1.4) node { $H_0 + H_1$ };
\draw[] (-1, 0.7) node { $H_1 + H_2$ };
\draw[] (-1, 0) node { $H_2 + H_0$ };

\draw[thick] (0, 3.5) -- (0.7, 3.5);
\draw[thick] (1.4, 3.5) -- (2.1, 3.5);
\draw[thick] (2.8, 3.5) -- (3.5, 3.5);
\draw[thick] (4.2, 3.5) -- (4.9, 3.5);

\draw[thick] (0.7, 2.8) -- (1.4, 2.8);
\draw[thick] (2.1, 2.8) -- (2.8, 2.8);
\draw[thick] (3.5, 2.8) -- (4.2, 2.8);

\draw[thick] (0.7, 2.1) .. controls (1.6, 2.5) and (1.9, 2.5) .. (2.8, 2.1);
\draw[thick] (2.1, 2.1) .. controls (3, 1.7) and (3.3, 1.7) .. (4.2, 2.1);

\draw[thick] (0, 1.4) -- (0.7, 1.4);
\draw[thick] (1.4, 1.4) -- (2.1, 1.4);
\draw[thick] (2.8, 1.4) -- (3.5, 1.4);
\draw[thick] (4.2, 1.4) -- (4.9, 1.4);
\draw[thick] (0.7, 1.4) -- (1.4, 1.4);
\draw[thick] (2.1, 1.4) -- (2.8, 1.4);
\draw[thick] (3.5, 1.4) -- (4.2, 1.4);

\draw[thick] (0.7, 0.7) -- (1.4, 0.7);
\draw[thick] (2.1, 0.7) -- (2.8, 0.7);
\draw[thick] (3.5, 0.7) -- (4.2, 0.7);
\draw[thick] (0.7, 0.7) .. controls (1.6, 1.1) and (1.9, 1.1) .. (2.8, 0.7);
\draw[thick] (2.1, 0.7) .. controls (3, 0.3) and (3.3, 0.3) .. (4.2, 0.7);

\draw[thick] (0.7, 0) .. controls (1.6, 0.4) and (1.9, 0.4) .. (2.8, 0);
\draw[thick] (2.1, 0) .. controls (3, -0.4) and (3.3, -0.4) .. (4.2, 0);
\draw[thick] (0, 0) -- (0.7, 0);
\draw[thick] (1.4, 0) -- (2.1, 0);
\draw[thick] (2.8, 0) -- (3.5, 0);
\draw[thick] (4.2, 0) -- (4.9, 0);

\foreach \h in {0, 0.7, 1.4, 2.1, 2.8, 3.5}{
	\foreach \x in {0, 1.4, 2.8, 4.2} {
		\draw[draw=black, fill=white, thick] (\x, \h) circle (3pt);
	}
	\foreach \y in {0.7, 2.1, 3.5, 4.9}{
		\filldraw (\y, \h) circle (3pt);
	}
}
\end{tikzpicture}
\caption{\textbf{Representative Hamiltonians for gapped phases and quantum critical points in $H = g_2 H_2 - g_1 H_1 + g_0 H_0$}.
Fermionic chains of $L =4$ unit cells are displayed, each comprising two sublattice sites denoted by empty and filled circles.
The unpaired Majorana fermions at the edges highlight the topological nature of $H_1$, $H_2$, and $H_1 + H_2$.
Note that $H_1 + H_2$ has the topologically protected edge modes although its bulk is gapless.}
\label{fig: chain}
\end{figure}

Let us consider a noninteracting $\alpha$-neighbor Hamiltonian for a one-dimensional chain of Majorana fermions (so-called the $\alpha$-chain) \cite{Ruben_topology_fermion}
\begin{align}
H_{\alpha} = -\frac{i}{2} \sum_{n} c_{2n} c_{2(n+\alpha)-1} \equiv  \frac{i}{2} \sum_{n} a_{n+\alpha}b_{n} ,
\label{eq: alpha_chain}
\end{align}
where $a_{n} \equiv c_{2n-1}$ and $b_{n} \equiv c_{2n}$ denote Majorana fermions at odd and even sublattice sites, respectively (marked as the blank and filled circles in Fig.~\ref{fig: chain}).
Each unit cell consists of Majorana fermions $a_n$ and $b_n$.
Then, the free fermion model in Eq.~(\ref{eq: free_fermion}) can be compactly written as $H = g_2 H_2 - g_1 H_1 + g_0 H_0$.
Without loss of generality, we let $g_0 + g_1 + g_2 = 4$.

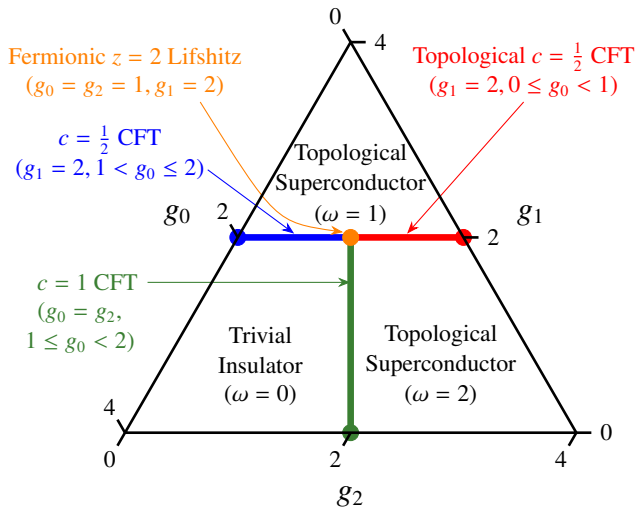
\begin{figure}[t]
\usetikzlibrary {arrows.meta}
\begin{tikzpicture}[scale=1, >=Stealth, domain=0:6]
\draw[] (0.7, {1.5 * sqrt(3) + 0.3}) node {\large $g_0$ };
\draw[] (3, {-0.2 * sqrt(3)-0.5}) node {\large $g_2$ };
\draw[] (5.4, {1.5 * sqrt(3) + 0.3}) node {\large $g_1$ };

\draw[thick, color = blue, line width = 2.5] (60:3) -- +(1.5,0);
\draw[thick, color = red, line width = 2.5] (3, {3*sqrt(3)/2}) -- +(1.5,0);
\draw[thick, color = OliveGreen, line width = 2.5] (3,0) -- (3, {3*sqrt(3)/2});

\draw[draw=orange, fill=orange] (3, {3*sqrt(3)/2}) circle (3pt);
\draw[draw=blue, fill=blue] (60:3) circle (3pt);
\draw[draw=red, fill=red] (4.5, {3*sqrt(3)/2}) circle (3pt);
\draw[draw=OliveGreen, fill=OliveGreen] (3,0) circle (3pt);

\draw[] (1.8, {3*sqrt(3)/4}) node { Trivial };
\draw[] (1.8, {3*sqrt(3)/4-0.4}) node {Insulator};
\draw[] (1.8, {3*sqrt(3)/4-0.8}) node {($\omega = 0$)};

\draw[] (4.2, {3*sqrt(3)/4})  node { Topological };
\draw[] (4.2, {3*sqrt(3)/4 - 0.4})  node { Superconductor };
\draw[] (4.2, {3*sqrt(3)/4 - 0.8})  node { ($\omega = 2$) };

\draw[] (3, 3.7) node { Topological };
\draw[] (3, 3.3) node {Superconductor};
\draw[] (3, 2.9) node {($\omega = 1$)};

\draw[color=blue] (-0.2, 3.9) node { $c=\frac{1}{2}$ CFT };
\draw[color=blue] (-0.2, 3.5) node {$(g_1 = 2, 1 < g_0 \leq 2)$};
\draw[->, color=blue] (1.1, {3*sqrt(3)/2 + 0.9}) -- (2.25, {3*sqrt(3)/2 + 0.05});

\draw[color=red] (5.3, 5) node {Topological $c=\frac{1}{2}$ CFT };
\draw[color=red] (5.3, 4.6) node {$(g_1 = 2, 0\leq g_0 < 1)$};
\draw[->, color=red] {(5.3, 4.4)} -- (3.75, {3*sqrt(3)/2 + 0.05});

\draw[color=OliveGreen] (-0.5, 2) node { $c=1$ CFT };
\draw[color=OliveGreen] (-0.6, 1.6) node {$(g_0 = g_2,$};
\draw[color=OliveGreen] (-0.6, 1.2) node {$1\leq g_0 < 2)$};
\draw[->, color=OliveGreen] (0.3,2) -- (3, 2);

\draw[color=orange] (0, 5) node {Fermionic $z=2$ Lifshitz};
\draw[color=orange] (0, 4.6) node {$(g_0= g_2 = 1, g_1 = 2)$};
\draw[->, color=orange, smooth] (1, 4.4) .. controls (2, 2.8) .. (2.9, {1.5 * sqrt(3) + 0.07});


\draw[line width = 1] (0,0) -- (60:6) -- (6,0) -- (0,0);
\draw[line width = 1] (0,0) -- (-120:0.2);
\draw[] (-120:0.4) node {0};
\draw[line width = 1] (3,0) -- +(-120:0.2);
\draw[] (3,0) + (-120:0.4) node {2};
\draw[line width = 1] (6,0) -- +(-120:0.2);
\draw[] (6,0) + (-120:0.4) node {4};

\draw[line width = 1] (6,0) -- +(0:0.2);
\draw[] (6,0) + (0:0.4) node {0};
\draw[line width = 1] ($(6,0) + (120:3)$) -- +(0:0.2);
\draw[] ($(6,0) + (120:3) + (0:0.4)$) node {2};
\draw[line width = 1] (60:6) -- +(0:0.2);
\draw[] ($(60:6) + (0:0.4)$) node {4};

\draw[line width = 1] (60:6) -- +(120:0.2);
\draw[] (60:6) + (120:0.4) node {0};
\draw[line width = 1] (60:3) -- +(120:0.2);
\draw[] (60:3) + (120:0.4) node {2};
\draw[line width = 1] (0,0) -- +(120:0.2);
\draw[] (120:0.4) node {4};

\end{tikzpicture}
\caption{\textbf{Ternary phase diagram of $H = g_2 H_2 - g_1 H_1 + g_0 H_0$ with $g_0+g_1+g_2 = 4$.}
The integer $\omega$ is the topological invariant for the gapped phases.
At the phase boundaries (colored solid lines), quantum critical states are described by CFTs with central charge $c$.
The representative Hamiltonians for these critical states are denoted by blue, red, and green circles, representing $2(H_0-H_2)$, $2(H_2-H_1)$, and $2(H_1+H_2)$, respectively.
The tricritical point (orange circle) is the only nonconformal critical point described by the fermionic Lifshitz theory with the dynamical critical exponent $z=2$.
It serves as the QCP for the continuous phase transition between the two topologically distinct $c=\frac{1}{2}$ CFTs (red and blue lines).}
\label{fig: phases}
\end{figure}

The $\alpha$-chain respects the particle-hole symmetry $\mathcal{C}$ and the time-reversal symmetry $\mathcal{T}$, which are characterized by the following relations: 
\begin{align}
\mathcal{T}H_{\alpha} \mathcal{T}^{-1} &= H_{\alpha},
\\
\mathcal{C}H_{\alpha} \mathcal{C}^{-1} &= - H_{\alpha}.
\label{eq: TC}
\end{align}
The symmetries are governed by antiunitary operators that transform Majorana operators as follows:
\begin{align}
\mathcal{C} c_j \mathcal{C}^{-1} &= c_j \label{eq: C},
\\
\mathcal{T}c_j\mathcal{T}^{-1} &= (-1)^j c_j \label{eq: T}.
\end{align}
Notably, the square of both symmetry operators satisfy $\mathcal{T}^2 = \mathcal{C}^2 = +1$ for the $\alpha$-chain.
Hence, our model $H = g_2 H_2 - g_1 H_1 + g_0 H_0$ belongs to the BDI symmetry class, allowing for various SPT phases and QCPs \cite{Kitaev_10foldway, Ryu_10foldway, Chiu_SPTclassify}.
Specifically, it encompasses three different gapped phases characterized by the integer topological invariant $\omega = 0, 1, 2$, three quantum critical states described by CFTs, and a nonconformal tricritical point situated at the intersection of the three conformal critical lines (see Fig.~\ref{fig: phases}) \cite{Ruben_topology_fermion, Ruben_1dSPT, Ruben_SETcriticality}.

To understand the available quantum phases of the model, let us first examine the $g_2 = 0$ limit, which corresponds to the left edge of the ternary diagram in Fig.~\ref{fig: phases}.
In this limit, the Hamiltonian $H$ simplifies to the well-known Kitaev chain: $H = g_0  H_0 - (4-g_0) H_1$ \cite{Kitaev_pwave}.
Its ground state transitions from a trivial insulator ($\omega = 0$) to a topological $p$-wave superconductor ($\omega = 1$) through a continuous phase transition at $g_0 = g_1 = 2$ (blue circle in Fig.~\ref{fig: phases}).
The fixed-point Hamiltonians for the trivial and topological phases of the Kitaev model are represented by the 0-chain ($H_0$) and the 1-chain ($H_1$), respectively.
The unpaired Majorana fermions at each end of the 1-chain exemplify the bulk-boundary correspondence for the topological superconductor (see Fig.~\ref{fig: chain}).
At the critical point $g_0 = g_1 = 2$, the bulk gap smoothly closes, and the low-energy degrees of freedom are linearly dispersing Majorana fermions.
Thus, the QCP is described by the Ising CFT with a central charge $c=1/2$.

In the $g_1 = 0$ limit (bottom edge in Fig.~\ref{fig: phases}), another phase transition exists between a trivial insulator and another topological superconductor, with $H_2$ as its fixed-point Hamiltonian. 
Note that the 2-chain of length $2L$ is effectively two copies of the 1-chain of length $L$ (Fig.~\ref{fig: chain}), implying that $H_2$ describes a topological superconductors with a winding number $\omega = 2$.
At the critical point $g_0 = g_2 = 2$ (marked as the green circle in Fig.~\ref{fig: phases}), the system features two gapless Majorana fermions in the bulk, leading to a critical theory characterized by the $c=1$ Gaussian CFT.

Furthermore, the phase diagram includes another critical theory representing the continuous phase transition between topological superconductors with distinct topological invariants $\omega = 1$ and $\omega = 2$.
Let us focus on the $g_0 = 0$ limit (right edge in Fig.~\ref{fig: phases}).
The critical Hamiltonian $H^{(L)} \propto -H_1^{(L)} + H_2^{(L)}$ for a chain of length $L$ is equal to the critical Hamiltonian for the Kitaev chain $H_0^{(L-1)} - H_1^{(L-1)}$ of length $(L-1)$, plus unpaired Majorana fermions at each end of the chain (Fig.~\ref{fig: chain}). 
Therefore, the critical theory must also be the $c=1/2$ Ising CFT, but with topologically protected Majorana edge modes \cite{Ruben_topology_fermion}.

Given that the critical phase boundary between the $\omega = 1$ and $\omega = 2$ states (red line in Fig.~\ref{fig: phases}) exhibits nontrivial topological features, this topological Ising CFT must be distinguished from the conventional Ising CFT that describes quantum phase transitions between the $\omega = 0$ and $\omega = 1$ states (blue line in Fig.~\ref{fig: phases}).
Hence, along the cut with $g_1 = 2$ (blue and red lines in Fig.~\ref{fig: phases}), a nonconformal quantum phase transition emerges (orange circle in Fig.~\ref{fig: phases}), bridging two CFTs with the same central charge $c=1/2$ but differing topological properties \cite{Ruben_topology_fermion}. 
As the linearly dispersing Majorana fermion becomes quadratically dispersing at this phase transition between two gapless states, the tricritical point at $(g_0, g_1, g_2) = (1, 2, 1)$ is nonconformal and is described by the fermionic Lifshitz theory with a dynamical critical exponent $z=2$.

\section{Entanglement Negativity of Fermionic Gaussian States \label{sec: negativity}}

This section reviews how to define and calculate the entanglement negativity for many fermion systems.
In Sec.~\ref{ssec: mixed_state_entanglement}, we first discuss the separability of a generic density operator \cite{Peres_separability, Horodecki_separability} and motivate the conventional definition of entanglement negativity from the necessary condition for a quantum state to be separable \cite{Vidal_negativity, negativity_monotone}.
Then, Sec.~\ref{ssec: fermionic_partial_transpose} elaborates on two unfavorable properties of the conventional entanglement negativity and reviews the axiomatic redefinition of the \emph{fermionic} entanglement negativity to cure all those unfavorable features \cite{Shapourian_partial_TR, Shapourian_fermion_negativity, Shiozaki_fSPT_invariant, Eisler_partial_transpose, Shapourian_fSPT_invariant}.
At last, we explain how to calculate the fermion version of entanglement negativity for a Gaussian mixed state using the two-point correlation functions in Sec.~\ref{ssec: free_fermion_negativity} \cite{Peschel_RDMcorrelation, negativity_bound, Fagotti_XYdisjoint, Shapourian_finiteTnegativity, twisted_negativity}.

\subsection{Entanglement of a mixed quantum state \label{ssec: mixed_state_entanglement}}

When a pure state $|\Psi\rangle \in \mathcal{H} = \mathcal{H}_A \otimes \mathcal{H}_B$ is prepared for a composite system $A\cup B$, the two subsystems $A$ and $B$ are said to be entangled if the state $|\Psi\rangle$ cannot be decomposed into any product state $|\psi_A \rangle \otimes |\psi_B\rangle$, where the quantum states $|\psi_A\rangle$ and $|\psi_B\rangle$ are prepared in the local Hilbert spaces $\mathcal{H}_A$ and $\mathcal{H}_B$, respectively.
The definition can be similarly extended to a mixed state; a density operator $\rho$ is said to be separable between the two subsystems $A$ and $B$ if there exists a convex combination
\begin{align}
\rho = \sum_{\alpha} p_\alpha \, \rho^{(\alpha)}_A \otimes \rho^{(\alpha)}_B,
\end{align}
where $p_\alpha \geq 0$, $\sum_{\alpha} p_\alpha = 1$, and $\rho^{(\alpha)}_A$ and $\rho^{(\alpha)}_B$ are the density operators for the subsystems $A$ and $B$, respectively.
The separable state has only classical correlations and no quantum entanglement because it can be prepared by local operations and classical communications (LOCC).
Therefore, an entangled state must be nonseparable.

In general, determining the separability of a generic mixed state is proven to be an NP-hard problem \cite{Gurvits_NPhard, Gharibian_NPhard}.
However, a computable diagnostic exists for the \emph{non}separability of a quantum state based on the positive partial transpose (PPT) criterion; the partial transpose of a separable state has only nonnegative eigenvalues \cite{Peres_separability, Horodecki_separability}.
When a quantum system is partitioned into two parts $A$ and $B$, the partial transpose of the density operator $\rho$ is defined as
\begin{align}
\rho^{T_A} &=\sum_{m,n;\mu,\nu} \rho_{m\mu,n\nu} \left[ \left |e^{(A)}_m,e^{(B)}_{\mu} \right \rangle \left \langle e^{(A)}_n, e^{(B)}_{\nu} \right |\right]^{T_A}
\nonumber \\
&\equiv \sum_{m,n;\mu,\nu} \rho_{m\mu,n\nu} \left |e^{(A)}_n,e^{(B)}_{\mu} \right\rangle \left \langle e^{(A)}_m, e^{(B)}_{\nu} \right |,
\label{eq: boson_PT}
\end{align}
where $\left \{ \left | e^{(A)}_m, e^{(B)}_{\mu} \right \rangle \right \}$ is the local basis of the Hilbert space $\mathcal{H}_A \otimes \mathcal{H}_B$.
If a quantum state $\rho$ is separable, its partial transpose
\begin{align}
\rho^{T_A} &=  \sum_\alpha p_\alpha \left( \rho_A^{(\alpha)} \right)^{T} \otimes \rho_B^{(\alpha)}
\end{align}
is a convex combination of positive semidefinite operators $\left( \rho_A^{(\alpha)} \right)^{T}$ and $\rho_B^{(\alpha)}$.
Thus, $\rho^{T_A}$ can be considered a separable quantum state having nonnegative eigenvalues.
On the contrary, the partial transpose of a generic nonseparable state is not necessarily positive semidefinite.
Hence, negative eigenvalues of the partially transposed density operator $\rho^{T_A}$ imply that the quantum state $\rho$ is not separable \cite{Peres_separability, Horodecki_separability}.

Then, we can define an entanglement measure, the so-called logarithmic entanglement negativity (or the negativity in short):
\begin{align}
\mathcal{E}_b = \ln \left \lvert \, \rho^{T_A} \right \rvert \equiv \ln \left( \sum_{n} \left| \lambda_n \right| \right),
\label{eq: bosonic_negativity}
\end{align}
where the norm $\left \lvert \, \cdot \, \right \rvert$ is defined as a sum of absolute values of the eigenvalues $\lambda_n$.
It quantifies a mixed quantum state's violation of the PPT criterion \cite{Vidal_negativity, Peres_separability, Horodecki_separability}.
While the vanishing negativity does not guarantee the separability of a quantum state, the nonzero negativity implies finite quantum entanglement between the two subsystems $A$ and $B$.
Furthermore, the negativity is a mixed state entanglement monotone, i.e., LOCCs cannot increase the negativity \cite{Vidal_negativity}.

\subsection{Partial transpose and entanglement negativity for fermions \label{ssec: fermionic_partial_transpose}}

Although the previous definition of the partial transpose [Eq.~(\ref{eq: boson_PT})] works well for bosons and spin systems, it has some unfavorable properties for fermionic systems due to the anticommuting exchange statistics \cite{Eisler_partial_transpose, Shapourian_fSPT_invariant, Shiozaki_fSPT_invariant, Shapourian_partial_TR, Shapourian_fermion_negativity}: $\left ( \rho^{T_A} \right)^{T_B} \neq \rho^T$, and the partial transpose does not preserve the tensor product structure for $n$ independent flavors of fermions, i.e.,
\begin{align}
\left( \rho_1 \otimes  \rho_2 \otimes \cdots \otimes \rho_n  \right )^{T_A} &\neq \rho_1^{T_A} \otimes \rho_2^{T_A} \otimes \cdots \otimes \rho_n^{T_A},
\label{eq: otimes_no_commute_PT}
\end{align}
which implies
\begin{align}
\mathcal{E}_b \left( \rho_1 \otimes \cdots \otimes \rho_n \right) \neq \mathcal{E}_b (\rho_1) + \cdots + \mathcal{E}_b (\rho_n).
\end{align}
Hence, a different definition of partial transpose is necessary for fermions to cure these ``unnatural'' properties.

For a physical state $\rho$ preserving fermion parity, one can axiomatically define the \emph{fermionic} partial transpose $\rho^{R_A}$ by demanding several physically natural conditions \cite{Shiozaki_fSPT_invariant, Shapourian_fermion_negativity}:
\begin{enumerate}
\item Be consistent with the full transpose: $\left ( \rho^{R_A} \right)^{R_B} = \rho^T$,
\item Preserve the identity: $\mathbb{I}^{R_A} = \mathbb{I}$,
\item Commute with the block canonical transformations: when a system of $N = N_A + N_B$ Majorana fermions are partitioned into two subsystems $A$ and $B$, $\left(U\rho U^\dagger\right)^{R_A} = U \rho^{R_A} U^\dagger$ under a canonical transformation $U$ which acts on the two subsystems independently, i.e., $U$ is a unitary transformation generating $O(N_A) \times O(N_B)$ rotations of Majorana fermions for the two subsystems.
\end{enumerate}
The third condition is necessary for the consistency of the fermionic partial transpose under the independent local change of bases of the two subsystems.
These three conditions and the conservation of fermion parity strongly constrain the analytical structure of the fermionic partial transpose.

Consider a general physical state for $N = N_A + N_B$ Majorana fermions
\begin{align}
\rho = \sum_{m=1}^{N_A} \sideset{}{'}\sum_{\substack {n=1\\ m+n = \mathrm{even}}}^{N_B} \rho_{p_1 \cdots p_{m}, q_1 \cdots q_{n}} c_{p_1} \cdots c_{p_m} \tilde{c}_{q_1} \cdots \tilde{c}_{q_n}, 
\label{eq: bipartition_general_rho}
\end{align}
where  $\{ c_1, \cdots, c_{N_A} \}$ and $\{ \tilde{c}_1, \cdots, \tilde{c}_{N_B} \}$ are two sets of Majorana fermions spanning the Fock spaces of the subsystems $A$ and $B$, respectively.
Since a density operator must be physical, the primed sum $\sum '$ in Eqs.~(\ref{eq: bipartition_general_rho}) and (\ref{eq: negativity_def}) is restricted to the terms with even fermion parity, i.e., $m+n$ must be even.
By the three conditions mentioned earlier, the fermionic partial transpose with respect to the subsystem $A$ introduces an additional multiplicative factor of $i^m$ to each term $c_{p_1} \cdots c_{p_m} \tilde{c}_{q_1} \cdots \tilde{c}_{q_n}$, where $m$ signifies the count of Majorana fermions within the subsystem $A$ \cite{Shapourian_partial_TR, Shapourian_fermion_negativity, Shiozaki_fSPT_invariant}:
\begin{align}
\rho^{R_A} \equiv  \sum_{m=1}^{N_A} \sideset{}{'}\sum_{\substack {n=1\\ m+n = \mathrm{even}}}^{N_B} i^m \rho_{p_1 \cdots p_{m}, q_1 \cdots q_{n}} c_{p_1} \cdots c_{p_m} \tilde{c}_{q_1} \cdots \tilde{c}_{q_n},
\label{eq: negativity_def}
\end{align}

Unlike the conventional partial transpose $(\, \cdot\, )^{T_{A}}$, one can show that the fermionic partial transpose preserves the tensor product structure \cite{Shapourian_partial_TR, Shapourian_fermion_negativity, Shiozaki_fSPT_invariant}
\begin{align}
\left( \rho_1 \otimes  \rho_2 \otimes \cdots \otimes \rho_n  \right )^{R_A} &= \rho_1^{R_A} \otimes \rho_2^{R_A} \otimes \cdots \otimes \rho_n^{R_A}.
\label{eq: fermionic_transpose_tensor}
\end{align}
It does not preserve Hermiticity.
Instead, the fermionic partial transpose $\rho^{R_A}$ is pseudo-Hermitian, i.e.,
\begin{align}
\left( \rho^{R_A} \right)^\dagger = (-1)^{F_A} \rho^{R_A} (-1)^{F_A},
\label{eq: pseudo-Hermiticity}
\end{align}
where $(-1)^{F_A} = \prod_{n \in A} (i c_{2n-1} c_{2n})$ measures the fermion parity of the subsystem $A$ \cite{twisted_negativity}.
Since the fermionic partial transpose of a density operator no longer has real eigenvalues, we cannot define the fermion version of the entanglement negativity as a measure of the PPT criterion violation.
However, we can still define the fermionic logarithmic negativity with a standard generalization of the norm of a linear operator \cite{Shapourian_partial_TR, Shapourian_fermion_negativity}:
\begin{align}
\mathcal{E}_f = \ln \left \lVert \, \rho^{R_A} \right \rVert \equiv \ln \mathrm{Tr} \left(\sqrt{\rho^{R_A} \left( \rho^{R_A} \right)^\dagger } \, \right).
\label{eq: log_fermion_negativity}
\end{align}
Due to Eq.~(\ref{eq: fermionic_transpose_tensor}), the fermioic negativity is additive for $n$ independent flavors of fermions, i.e.,
\begin{align}
\mathcal{E}_f \left( \rho_1 \otimes \cdots \otimes \rho_n \right) = \mathcal{E}_f (\rho_1) + \cdots + \mathcal{E}_f (\rho_n).
\label{eq: additive}
\end{align}

Even though we cannot directly relate the fermionic (logarithmic) negativity to the PPT criterion, the fermionic negativity preserves many important properties of the conventional entanglement negativity.
In particular, the fermionic negativity is an entanglement monotone, i.e., it does not increase under LOCCs such as appending unentangled ancilla, local projective measurements, and tracing out an entangled ancilla.
Furthermore, it is zero for a separable state of fermions although the vanishing fermionic negativity does not guarantee the separability of a density operator \cite{Shapourian_fermion_negativity}.

\subsection{Entanglement negativity of noninteracting fermions \label{ssec: free_fermion_negativity}}

The previous two subsections discuss the partial transpose and the entanglement negativity for a general quantum state of bosons and fermions.
This subsection provides a practical summary of calculating the logarithmic negativity for a mixed quantum state of noninteracting Majorana fermions.

\subsubsection{Entanglement Hamiltonian from covariance matrix}
Let us consider a Gaussian state for $2N$ Majorana fermions
\begin{align}
\rho = \frac{1}{Z} e^{-\mathcal{H}}
\end{align}
with a quadratic entanglement Hamiltonian
\begin{align}
\mathcal{H} = \frac{i}{4} \sum_{j,k=1}^{2N} A_{jk} c_j c_k,
\end{align}
where $A_{jk} = -A_{kj} \in \mathbb{R}$, and $c_j$ are Majorana fermions.
The complete information about the Gaussian state is encoded in the two-point correlation functions $\langle i c_j c_k \rangle$ because higher-point correlation functions can be calculated using Wick's theorem.
Thus, we can calculate the entanglement negativity from the covariance matrix of Majorana fermions \cite{Peschel_RDMcorrelation, Shapourian_finiteTnegativity}
\begin{align}
\gamma_{jk} &= \frac{i}{2} \langle [ c_j , c_k ] \rangle =
\begin{cases}
~0, &j = k
\nonumber \\
~\langle i c_j c_k \rangle= \mathrm{Tr} \left(e^{-\mathcal{H}} i c_j c_k \right) / Z, &j\neq k
\end{cases}.
\end{align}

To calculate the covariance matrix, we first diagonalize the entanglement Hamiltonian by block factorization of the real skew-symmetric matrix $A$
\begin{align}
A = Q\Sigma Q^T = Q \left[ \bigoplus_{n=1}^{N}
\begin{pmatrix}
0 & \lambda_n
\\
-\lambda_n & 0
\end{pmatrix}
\right]Q^T,
\label{eq: block_factorize_skewsymmetric}
\end{align}
where $\lambda_n \geq 0$ are the nonnegative eigenvalues of $iA$, and $Q \in O(2N, \mathbb{R})$ is a real orthogonal matrix.
Hence,
\begin{align}
\mathcal{H} &= \frac{i}{4} \sum_{n=1}^{N}
\begin{pmatrix}
d_n' & d_n''
\end{pmatrix}
\begin{pmatrix}
0 & \lambda_n
\\
-\lambda_n & 0
\end{pmatrix}
\begin{pmatrix}
d_n'
\\
d_n''
\end{pmatrix}
= \frac{i}{2} \sum_{n=1}^N \lambda_n d_n' d_n''
\\
&= \sum_{n=1}^N \lambda_n \left( f_n^\dagger f_n - \frac{1}{2} \right),
\end{align}
where $d_n' = \sum_{j=1}^{2N} c_j Q_{j, 2n-1} = f_n + f_n^\dagger$ and $d_n'' = \sum_{j=1}^{2N} c_j Q_{j, 2n} = -i\left( f_n - f_n^\dagger \right)$ are the two Majorana normal modes, and $f_n^\dagger$ and $f_n$ are the fermionic creation and annihilation operators of the eigenstates.
Then, we can derive a general relationship between the covariance matrix $\gamma$ and the entanglement Hamiltonian $\mathcal{H}$:
\begin{align}
\gamma_{jk}
&=\sum_{n=1}^{N} \left( Q_{j,2n-1} Q_{k,2n} - Q_{j,2n}Q_{k,2n-1} \right)\left \langle i d_n' d_n'' \right \rangle
\label{eq: covariance_Majorana_mode}
\\
&=\left(
Q\left[
\bigoplus_{n=1}^N
\begin{pmatrix}
0 & -\tanh\left(\lambda_n / 2\right) 
\\
\tanh\left(\lambda_n / 2 \right) & 0
\end{pmatrix}
\right]Q^T
\right)_{jk}
\label{eq: gamma_factorize}
\\
\Rightarrow i\gamma &= \tanh\left( iA / 2 \right)
\Leftrightarrow e^{-iA} = \left(\mathbb{I} - i\gamma\right)\left(\mathbb{I} + i\gamma\right)^{-1}.
\label{eq: expiA}
\end{align}
In the second equality, we used $\left \langle i d_n' d_n'' \right \rangle = 2\left \langle f_n^\dagger f_n \right\rangle - 1$ and $\left \langle f_n^\dagger f_n \right\rangle = \left( e^{\lambda_n} +1 \right)^{-1}$.

\subsubsection{Fermionic partial transpose of a Gaussian state}
An essential feature of the fermionic partial transpose is the preservation of the Gaussian nature, i.e., the fermionic partial transpose of a Gaussian state is also Gaussian \cite{Shapourian_partial_TR}.
Let us denote $\rho^{R_A} \equiv \rho_+$ and $\left(\rho^{R_A}\right)^\dagger \equiv \rho_-$ to simplify the notation.
It can be shown from general observation that Eqs.~(\ref{eq: bipartition_general_rho}) and (\ref{eq: negativity_def}) imply
\begin{align}
\left \langle c_{j_1} \cdots c_{j_m} \tilde{c}_{k_1} \cdots \tilde{c}_{k_n} \right \rangle_+
&\equiv \mathrm{Tr} \left ( \rho_+ c_{j_1} \cdots c_{j_m} \tilde{c}_{k_1} \cdots \tilde{c}_{k_n} \right )
\\
&= i^m \, \mathrm{Tr} \left ( \rho c_{j_1} \cdots c_{j_m} \tilde{c}_{k_1} \cdots \tilde{c}_{k_n} \right )
\\
&\equiv i^m \left \langle c_{j_1} \cdots c_{j_m} \tilde{c}_{k_1} \cdots \tilde{c}_{k_n} \right \rangle
\label{eq: negativity_gaussian_proof}
\end{align}
for any correlation function of $m$ and $n$ Majorana fermions in the subsystems $A$ and $B$, respectively.
While Eq.~(\ref{eq: negativity_gaussian_proof}) is true for any quantum state $\rho$ and its fermionic partial transpose $\rho^{R_A}$, we can go further if $\rho$ is Gaussian:
\begin{align}
&\left \langle c_{j_1} \cdots c_{j_m} \tilde{c}_{k_1} \cdots \tilde{c}_{k_n} \right \rangle_+
= i^m \sum_{\substack{\textrm{all pairwise}\\ \textrm{contractions}}}
\left \langle c_{j_1} c_{j_2} \right \rangle\left \langle c_{j_3} c_{j_4} \right \rangle \cdots
\nonumber \\
&=i^m \sum_{\substack{\textrm{all pairwise}\\ \textrm{contractions}}}
\left \langle -i c_{j_1} c_{j_2} \right \rangle_{+} \left \langle -i c_{j_{3}} c_{j_4} \right \rangle_{+}
\cdots \Big \langle \tilde{c}_{k_{n-1}} \tilde{c}_{k_n} \Big \rangle_{+}  \pm  \cdots
\nonumber \\
&= \sum_{\substack{\textrm{all pairwise}\\ \textrm{contractions}}}
\left \langle c_{j_1} c_{j_2} \right \rangle_{+}\left \langle c_{j_3} c_{j_4} \right \rangle_{+} \cdots
 \Big \langle \tilde{c}_{k_{n-1}} \tilde{c}_{k_n} \Big \rangle_{+} \pm \cdots,
\end{align}
i.e., the correlation function $\langle \, \cdots \rangle_{+}$ with respect to $\rho^{R_A}$ also satisfies the Wick's theorem if $\rho$ is Gaussian.
Therefore, we can write $\rho_{\pm} = e^{-\mathcal{N^{\pm}}} / Z_{\pm}$ with non-Hermitian quadratic Hamiltonians
\begin{align}
\mathcal{N}^{\pm} = \frac{i}{4} \sum_{j,k=1}^{2N} W_{jk}^{\pm} c_j c_k,
\label{eq: negativity_Hamiltonian}
\end{align}
where $W^{\pm}_{jk} = -W^{\pm}_{kj} \in \mathbb{C}$ \cite{negativity_Hamiltonian}.

Using Eq.~(\ref{eq: negativity_gaussian_proof}), we can relate the covariance matrix $\gamma_{\pm}$ for $\rho_{\pm}$ to the covariance matrix $\gamma$ for $\rho$ \cite{negativity_bound, Shapourian_finiteTnegativity}:
\begin{align}
\left(\gamma_{\pm}\right)_{jk} &= \mathrm{Tr} \left(\rho_{\pm} i c_j c_k \right)
\nonumber \\
&=
\left \{
\begin{aligned}
\left(\pm i\right)^2 \langle ic_j c_k \rangle &= -\gamma_{jk}, ~~~ j, k \in A
\\
\pm i  \langle ic_j c_k \rangle &= \pm i\gamma_{jk}, ~~ \mathrm{only~one~of}~j, k \in A
\\
\langle ic_j c_k \rangle  &= \gamma_{jk}, \quad~\, j, k \in B
\end{aligned}
\right . .
\label{eq: gamma_pm}
\end{align}
Although $\rho_{\pm}$ are not Hermitian, their Gaussian nature permits the generalization of Eq.~(\ref{eq: expiA}) to the complex skew-symmetric diagonalizable matrix $W^{\pm}$ \cite{Fagotti_XYdisjoint}, i.e.,
\begin{align}
e^{-iW^{\pm}} = \left( \mathbb{I} - i\gamma_{\pm} \right) \left( \mathbb{I} + i\gamma_{\pm}\right)^{-1}.
\end{align}
Thus, the covariance matrix $\gamma$ of a Gaussian fermionic state $\rho$ is sufficient to determine the fermionic partial transpose $\rho^{R_A}$.

\subsubsection{Entanglement negativity from products of covariance matrices \label{ssec: product_covariance}}

To calculate the fermionic negativity $\mathcal{E}_f = \ln \mathrm{Tr} \sqrt{\rho_+ \rho_-}$, we need to compute $\rho_{+}\rho_{-}$.
Since
\begin{align}
\left[-\mathcal{N}^{+}, -\mathcal{N}^{-} \right] &=
\left[ \frac{1}{4} \sum_{jk} \left(-iW_{jk}^{+} \right)c_j c_k, \frac{1}{4} \sum_{lm} \left(-i W_{lm}^{-}\right) c_l c_m \right]
\nonumber \\
&= \frac{1}{4} \sum_{jk} \left[-iW^{+}, -iW^{-} \right]_{jk} c_j c_k,
\end{align}
the Baker-Campbell-Hausdorff (BCH) formula implies \cite{negativity_bound}
\begin{align}
\rho_{+}\rho_{-} &= \frac{1}{Z_{+}Z_{-}} e^{-\mathcal{N}^{+}}e^{-\mathcal{N}^{-}}
\nonumber \\
&= \frac{1}{Z_{+}Z_{-}} e^{-\mathcal{N}^{+} - \mathcal{N}^{-} + \frac{1}{2} \left[ -\mathcal{N}^{+}, -\mathcal{N}^{-} \right] +
\cdots}
\nonumber \\
&= \frac{1}{Z_{+}Z_{-}} e^{\frac{1}{4} \sum_{jk} \left( 
-iW^{+} -iW^{-} + \frac{1}{2} \left[-iW^{+}, -iW^{-} \right] +
\cdots
\right)_{jk} c_j c_k}
\nonumber \\
&\equiv \frac{1}{Z_+ Z_-} e^{-\frac{i}{4} \sum_{jk} W^{\times}_{jk} c_j c_k},
\end{align}
where $e^{-iW^{\times}} \equiv e^{-iW^{+}}e^{-iW^{-}}
= e^{-iW^{+} -iW^{-} + \frac{1}{2} \left[-iW^{+}, -iW^{-} \right] + \cdots}$.
Then, the product of Gaussian operators
\begin{align}
\rho_{+}\rho_{-} &= \frac{Z_{\times}}{Z_{+}Z_{-}} \frac{1}{Z_{\times}} e^{-\frac{i}{4} \sum_{jk} W^{\times}_{jk} c_j c_k}
\equiv \frac{Z_{\times}}{Z_{+}Z_{-}} \rho_{\times},
\label{eq: rho_times}
\end{align}
where $Z_{\sigma} = \mathrm{Tr} ~e^{-\frac{i}{4}\sum_{jk} W_{jk}^{\sigma} c_j c_k} = \mathrm{Pf}  \left(\mathbb{I} + e^{-iW^{\sigma}} \right)$ for $\sigma = +, -, \times$.
Then, the covariance matrix $\gamma_{\times}$ for $\rho_{\times}$ is calculated from $\gamma_{\pm}$:
\begin{align}
\frac{\mathbb{I} - i\gamma_{\times}}{\mathbb{I} + i\gamma_{\times}}
&= e^{-iW^{\times}} = e^{-iW^{+}}e^{-iW^{-}} = \frac{\mathbb{I} - i\gamma_{+}}{\mathbb{I} + i\gamma_{+}}\frac{\mathbb{I} - i\gamma_{-}}{\mathbb{I} + i\gamma_{-}}
\\
\Rightarrow \frac{\mathbb{I} + i\gamma_{\times}}{2}
&= \frac{\mathbb{I} + i\gamma_{-}}{2} \left( \frac{\mathbb{I} - \gamma_{+} \gamma_{-}}{2} \right)^{-1} \frac{\mathbb{I} + i\gamma_{+}}{2}.
\label{eq: gamma_x}
\end{align}
The fermionic negativity $\mathcal{E}_f$ can be calculated using the trace formulae for a fermionic Gaussian state (Appendix \ref{app: trace}):
\begin{align}
\mathcal{E}_f &= \ln \mathrm{Tr} \sqrt{\rho_{+}\rho_{-}}= \frac{1}{2} \ln \left( \frac{Z_{\times}}{Z_{+}Z_{-}} \right)
+ \ln \mathrm{Tr} \, \rho_{\times}^{1/2}
\\
&=\frac{1}{2} \ln \mathrm{Tr} \,\rho^2 + \ln \mathrm{Tr}\,\rho_{\times}^{1/2}
\equiv -\frac{1}{2}S_2 + \mathcal{E}_{\times}
\label{eq: negativity_formula}
\\
&=\frac{1}{4} \sum_{n=1}^{2N} \ln \left[ \left(1-\zeta_n\right)^2 + \zeta_n^2 \right]
\nonumber \\
&\qquad\qquad\qquad+ \frac{1}{2}\sum_{n=1}^{2N} \ln \left( \sqrt{1-\zeta^{\times}_n} + \sqrt{\zeta_n^{\times}}\, \right),
\label{eq: negativity_formula_eigenvalue}
\end{align}
where $\zeta_n$ and $\zeta^{\times}_n$ are the eigenvalues of $\left(\mathbb{I} + i\gamma\right)/2$ and $\left(\mathbb{I} + i\gamma_{\times}\right)/2$, respectively.
Note that the first term in Eq.~(\ref{eq: negativity_formula}) is nothing but the second R\'{e}nyi entropy $S_{2} =-\ln \mathrm{Tr}\,\rho^{2}$, which quantifies the purity of a quantum state $\rho$.
Since $\mathrm{Tr}\,\rho^2 \leq 1$, $\ln \mathrm{Tr}\, \rho^2 \leq 0$ penalizes the contribution from the entanglement between the total system $A\cup B$ and the environment such as a thermal reservoir.

\section{Entanglement Negativity of Adjacent Intervals \label{sec: adjacent}}

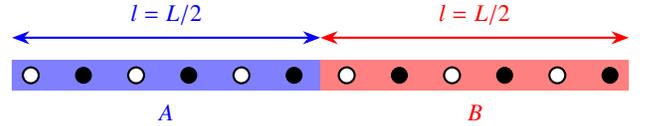
\begin{figure}[h]
\usetikzlibrary {arrows.meta}
\begin{tikzpicture}[>=Stealth]
\draw[blue] (1.8, 0.8) node { $l = L/2$ };
\draw [<->, thick, blue] (-0.25, 0.5) -- (3.85, 0.5);
\draw[red] (5.9, 0.8) node { $l = L/2$ };
\draw [<->, thick, red] (3.85, 0.5) -- (7.95, 0.5);
\draw[blue] (1.8, -0.5) node { $A$ };
\draw[red] (5.9, -0.5) node { $B$ };
\fill[semitransparent, blue] (-0.25,0.2) rectangle (3.85,-0.2);
\fill[semitransparent, red] (3.85,0.2) rectangle (7.95,-0.2);
\foreach \x in {0, 1.4, 2.8, 4.2, 5.6, 7} {
	\draw[draw=black, fill=white, thick] (\x, 0) circle (3pt);
	}
	
\foreach \y in {0.7, 2.1, 3.5, 4.9, 6.3, 7.7} {
	\filldraw (\y, 0) circle (3pt);
	}
\end{tikzpicture}
\caption{\textbf{Bipartition of the fermionic chain of length $L$ unit cells.}}
 \label{fig: bipartition}
\end{figure}

Let us consider an equal bipartition of the one-dimensional fermionic system of $L$ unit cells (Fig.~\ref{fig: bipartition}).
We calculate the fermionic negativity $\mathcal{E}_f$ of a thermal mixed state $\rho = e^{-H/T} / Z$, where the partition function $Z = \mathrm{Tr}\, e^{-H/T}$.
We consider both the open and periodic boundary conditions and assume an even number of the unit cells $L$.

\begin{figure*}[t]
\includegraphics[width=\textwidth]{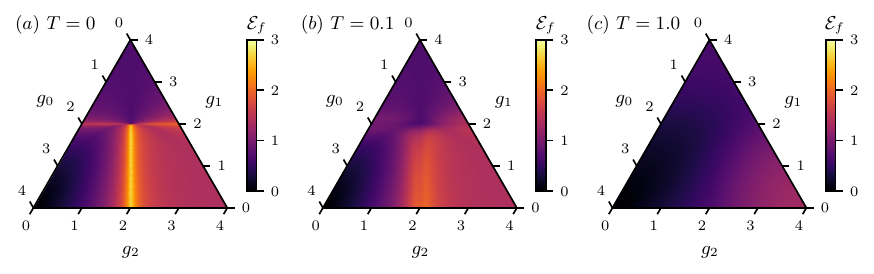}
\caption{\textbf{The fermionic negativity $\mathcal{E}_f$ of the two adjacent intervals of length $l = L / 2 = 50$ unit cells.}
Periodic boundary conditions are assumed.
(a) At $T=0$, the QCPs have large $\mathcal{E}_f$ and sharply distinguish three topologically distinct gapped phases.
(b) At $T = 0.1$, a faint double-peak structure near the critical lines appears.
While $\mathcal{E}_f$ near the critical points is significantly diminished, $\mathcal{E}_f$ of the gapped phases is not changed much.
(c) At $T = 1.0$, the sharp phase boundaries are no longer visible, and the high-temperature regime is entered.} 
\label{fig: bipartite_ternary} 
\end{figure*}

Following the method outlined in Sec.~\ref{ssec: free_fermion_negativity}, we map out the complete bipartite entanglement phase diagrams at $T\geq0$ (Fig.~\ref{fig: bipartite_ternary}).
The critical lines corresponding to CFTs display markedly elevated $\mathcal{E}_f$, resulting in a clear demarcation between the three topologically distinct gapped phases [depicted in Fig.~\ref{fig: bipartite_ternary} (a)].
Upon transitioning to $T>0$, the influence of thermal fluctuations becomes pronounced along the critical lines due to the dense energy spacing $\Delta < T$ relative to the temperature.
Although gaplessness bestows strong entanglement upon the quantum state at $T=0$, this characteristic also renders the state more vulnerable to decoherence at $T>0$.
However, the negativity $\mathcal{E}_f$ for the gapped phases remains relatively unaltered up to the crossover temperature $T_Q \sim \Delta$, as the substantial energy gap $\Delta > T$ serves to shield the ground state from significant thermal admixture with other excited states [as illustrated in Fig.~\ref{fig: bipartite_ternary} (b)].

The following subsections discuss more detailed analyses of $\mathcal{E}_f$ for $T\geq0$, concentrating on two distinct paths within the phase diagram.
In Sec.~\ref{ssec: gapped_QCP}, we explore the behavior of $\mathcal{E}_f(T)$ in proximity to the QCP that separates the two gapped phases, particularly along the trajectory $g_2 = 0$.
Turning attention to Sec.~\ref{ssec: gapless_QCP}, we delve into the examination of $\mathcal{E}_f(T)$ along the path $g_1 = 2$, elucidating how quantum entanglement portrays the fermionic $z=2$ Lifshitz transition between the topologically distinct quantum critical states.

\subsection{Gapped SPT phases and conformal critical point \label{ssec: gapped_QCP}}

\begin{figure}[t]
  \centering
  \includegraphics{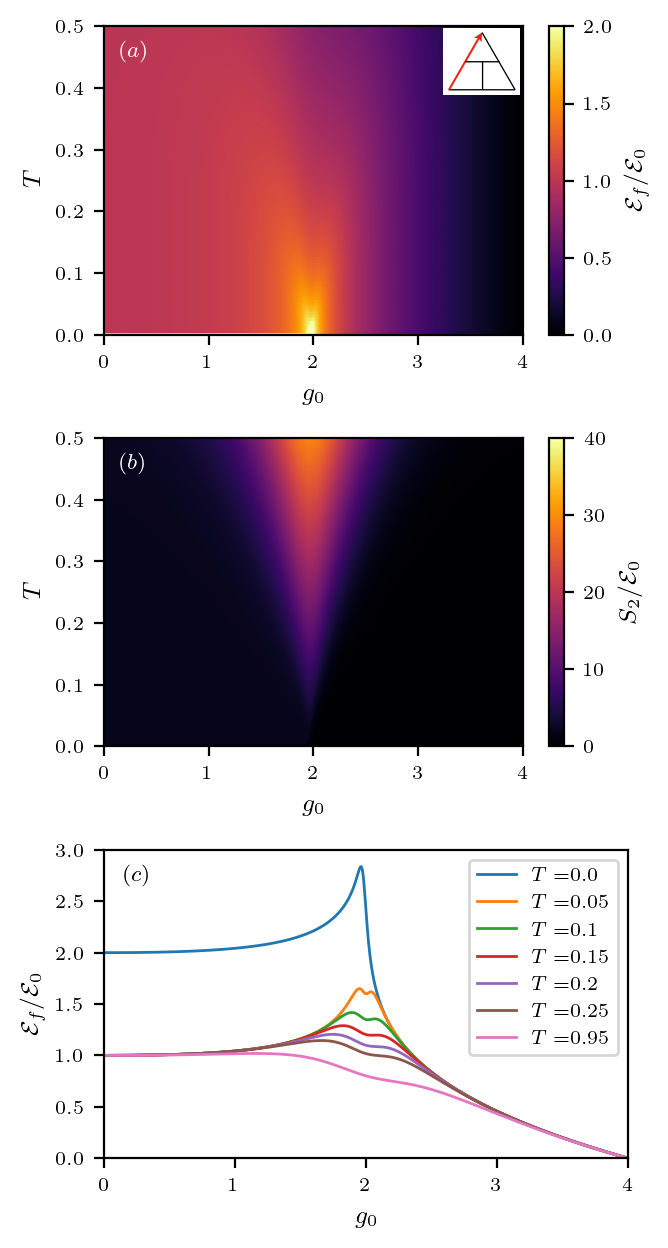}
  \caption{\textbf{Finite temperature entanglement negativity of Kitaev chain.} (a) The fermionic negativity $\mathcal{E}_f/\mathcal{E}_0$ and (b) the second R\'{e}nyi entropy $S_2/\mathcal{E}_0$ of the Kitaev chain, $H = g_0 H_0 - (4 - g_0) H_1$, for the two adjacent intervals of length $L / 2 = 50$ with the open boundary condition. $\mathcal{E}_0 = (\ln 2)/2$ is the fermionic negativity of the maximally entangled two Majorana zero modes. (c) The fermionic negativity $\mathcal{E}_f$ shows a double-peak structure at finite temperatures. The peaks are located at the crossover boundary of the quantum critical fan shown in (a).}
  \label{fig: Kitaev_fan}
\end{figure}
\begin{figure}[t]
  \centering
  \includegraphics{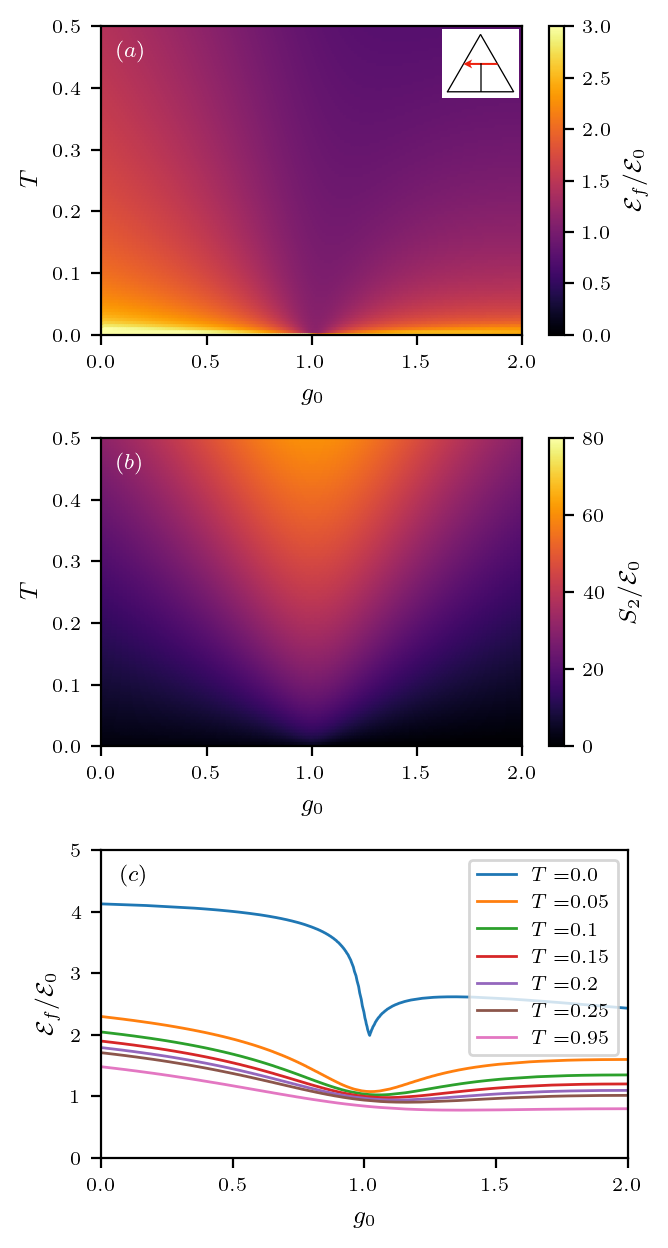}
  \caption{\textbf{Finite temperature entanglement negativity of Ising CFT.} (a) The fermionic negativity $\mathcal{E}_f/\mathcal{E}_0$ and (b) the second R\'{e}nyi entropy $S_2/\mathcal{E}_0$ of the Ising CFTs, $H = g_0 H_0 - 2 H_1 + (2-g_0) H_2$, for the two adjacent intervals of length $L / 2 = 50$ with the open boundary condition. (c) The fermionic negativity $\mathcal{E}_f$ shows a prominent dip at the Lifshitz tricritical point, $g_0 =1$.}
  \label{fig: CFT_fan}
\end{figure}

When $g_2 = 0$, our free fermion Hamiltonian is reduced to Kitaev's seminal model for the $p$-wave superconductor, $H = -(4-g_0) H_1 + g_0 H_0$ \cite{Kitaev_pwave}.
The quantum phase transition between the two gapped phases occurs at the QCP, $g_0 = g_c = 2$.
While the presence of nonzero temperature is a relevant perturbation at the QCP of the model, the intriguing quantum critical behavior does not abruptly disappear.
In the vicinity of the QCP at low temperatures, there is a region characterized by universal critical properties.
This region is commonly referred to as the quantum critical fan owing to its distinctive fan-shaped appearance within the phase diagram.
Near the QCP, the fermionic negativity $\mathcal{E}_f$ unveils the quantum critical fan at finite temperatures, whose crossover boundary can be seen as a sharp, V-shaped region within Fig.~\ref{fig: Kitaev_fan} (a).
The local maxima of $\mathcal{E}_f$ at $T>0$ are located at the crossover boundary of the quantum critical fan, as illustrated by the double peak structure in Fig.~\ref{fig: Kitaev_fan} (c).

For $g_0 \neq g_c = 2$, $\mathcal{E}_f$ displays weak temperature dependence outside the quantum critical fan, i.e., the temperature is below the energy gap, $T < \Delta$.
Fig.~\ref{fig: Kitaev_fan} (c) demonstrates that $\mathcal{E}_f$ remains nearly constant at different temperatures when the coupling strength is away from the QCP, i.e., $|g_0-g_c| \gg T$.
This behavior can be attributed to the exponential suppression of contributions from excited states, governed by the Gibbs factor $e^{-\Delta/T}$.
The excitation gap, denoted as $\Delta$, effectively preserves the ground state entanglement of SPTs low temperatures, $T < \Delta$.

Due to the topological nature of the $p$-wave superconductor ($0 \leq g_0 < 2$), $\mathcal{E}_f \sim \mathcal{E}_0 = (\ln 2)/2$ when $T>0$  [Fig.~\ref{fig: Kitaev_fan} (c)].
The entanglement mostly comes from the two neighboring Majorana fermions at the interface of the two partitions $A$ and $B$.
Hence, $\mathcal{E}_f$ is equal to the fermionic negativity of two maximally entangled Majorana zero modes $\mathcal{E}_0$ (Appendix \ref{app: E0}).
At $T = 0$, $\mathcal{E}_f \sim 2\mathcal{E}_0$ for the topological superconductor [the blue curve in Fig.~\ref{fig: Kitaev_fan} (c)] is twice larger than the finite temperature value $\mathcal{E}_f \sim \mathcal{E}_0 $ for open boundary conditions.
The long-distance entanglement between the left and right edge modes contributes additional entanglement $\mathcal{E}_{\mathrm{edge}} = \mathcal{E}_0$ (see Sec.~\ref{sec: disjoint}).
Although the finite temperature gapped phases are, strictly speaking, not SPTs with topologically protected edge modes \cite{finiteT_SPT_unstable}, they feature the signature of the SPT ground states up to the crossover temperature $T_Q \sim \Delta$ (Fig.~\ref{fig: bipartite_ternary}).

At $T>0$, thermal fluctuations immediately suppress the edge-to-edge entanglement $\mathcal{E}_0$. 
Since the energy cost $\varepsilon_1$ to excite the edge states is zero, arbitrarily weak thermal fluctuations can excite them.
Hence, the contribution of the two-fold degenerate Majorana zero modes to the thermal covariance matrix is
\begin{align}
\gamma_{2m-1.2n} &= U_{m,1}V_{n,1} \left( 2 \left\langle f_1^\dagger f_1 \right \rangle_T - 1\right)
\nonumber \\
&= U_{m,1}V_{n,1} \left( 2 \cdot \frac{1}{2} - 1 \right) = 0,
\end{align}
where $\left\langle f_1^\dagger f_1 \right \rangle_T = 1/\left(e^{\varepsilon_1 / T} + 1\right) = 1/2$, and $U$ and $V$ are orthogonal matrices for singular value decomposition of the hopping matrix $t$, whose definition can be found in Appendix \ref{app: covariance}.
Thus, $\mathcal{E}_{\mathrm{edge}}(T>0) = 0$, and $\mathcal{E}_f(T=0) - \mathcal{E}_f(T>0) = \mathcal{E}_0$ for any $T > 0$ [Fig.~\ref{fig: Kitaev_fan} (c)].

When temperatures exceed the energy gap, $T > \Delta$, the system crosses over to the quantum critical fan.
Due to the densely packed excitation spectrum relative to the temperature $T$, thermal fluctuations can make the quantum state highly mixed, thereby reducing its purity $\mathrm{Tr} \,\rho^2$.
Consequently, the fermionic negativity $\mathcal{E}_f = -\frac{1}{2}S_2 + \mathcal{E}_\times$ decreases, primarily due to the large second R\'{e}nyi entropy $S_2 = -\ln \mathrm{Tr}\,\rho^2$ [Fig.~\ref{fig: Kitaev_fan} (b)], originating from thermal fluctuations.
Thus, the fermionic negativity for two adjacent intervals undergoes a universal logarithmic decrease with temperature, $\mathcal{E}_f \sim -\ln T$.

The logarithmic temperature dependence of the negativity inside the quantum critical fan stems from the universal temperature scaling behavior of the correlation length within this region.
At $T=0$, the correlation length diverges at the QCP, resulting in macroscopically large entanglement $\mathcal{E}_f \sim \ln L$ \cite{Calabrese_QFTnegativity, Calabrese_negativity_extended, Shapourian_partial_TR}.
However, the correlation length remains finite as temperature increases, scaling as $\xi_T \sim 1/T$.
As a result, the bipartite entanglement between two blocks of length $L/2$ at $T>0$ can be viewed as the entanglement between two adjacent blocks of size $\xi_T$.
This yields $\mathcal{E}_f \sim \ln \xi_T$, which simplifies to $\mathcal{E}_f \sim -\ln T$.
The crossover boundary represents the point at which the energy gap $\Delta$ becomes comparable to the temperature $T$.
At this point, thermal fluctuations begin to play a pivotal role in reducing entanglement.
Therefore, substantial entanglement is not characteristic of states within the entire quantum critical fan but is most pronounced at the crossover boundary.

\subsection{Gapless SPT and fermionic Lifshitz criticalities \label{ssec: gapless_QCP}}

While the entanglement negativity of bosonic Lifshitz theory \cite{Lifshitz_boson, RK_separability} and the entanglement entropy of Lifshitz fermions \cite{Lifshitz_entropy, Wang_Lifshitz} have been extensively investigated, the fermionic negativity of the fermionic Lifshitz theory remains unexplored.
With $g_1 = 2$, we studied the fermionic negativity $\mathcal{E}_f$ along the critical phase boundary (the blue and red paths in Fig.~\ref{fig: phases}).
The Hamiltonian $H = g_0 H_0 - 2H_1 + (2-g_0) H_2$ is gapless for all $0 \leq g_0 \leq 2$, and there is the fermionic $z=2$ Lifshitz transition at $g_0 = g_c = 1$ (the orange circle in Fig.~\ref{fig: phases}) \cite{Ruben_topology_fermion}.
The fermionic negativity $\mathcal{E}_f$ shows a prominent dip at the fermionic Lifshitz tricritical point [Fig.~\ref{fig: CFT_fan} (a) and (c)].
Although the other gapless states ($g_0 \neq g_c$) have diverging correlation lengths, the ground state of the Lifshitz tricritical point has finite correlations only between two adjacent fermions, i.e., $\langle i a_m b_n \rangle = \delta_{m,n+1}$, when periodic boundary conditions are imposed.
Therefore, the fermionic negativity for the Lifshitz tricritical point $\mathcal{E}_f = \ln 2$ is constant regardless of the system size at $T = 0$ (Appendix \ref{app: Lifshitz}).
Since $\mathcal{E}_f \sim \ln L$ for the other gapless states represented by CFTs \cite{Calabrese_QFTnegativity, Calabrese_negativity_extended, Shapourian_partial_TR}, we can see the sudden drop of $\mathcal{E}_f$ at the tricritical point $g_0 = g_c = 1$ in Fig.~\ref{fig: CFT_fan} (c).

\begin{figure}[t]
\includegraphics{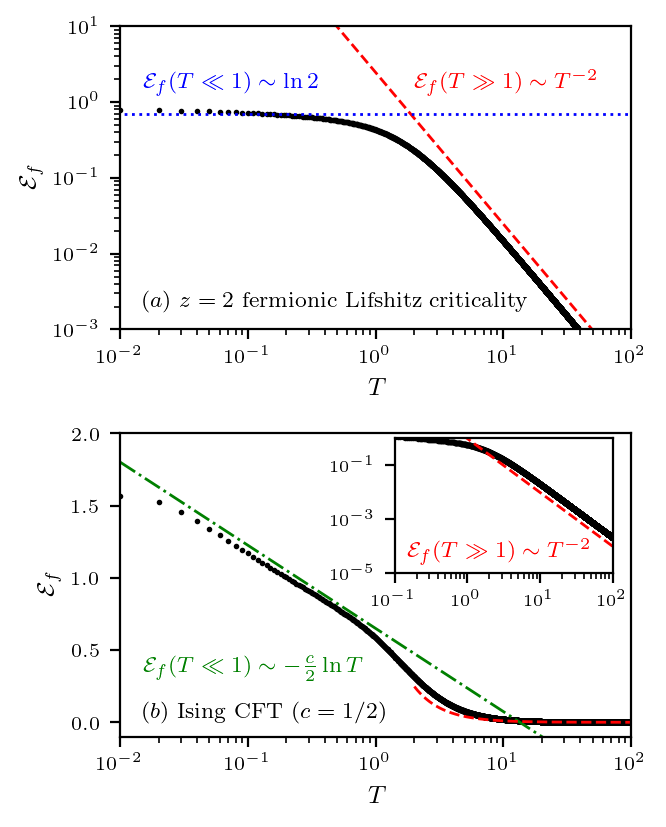}
\caption{\textbf{Bipartite fermionic negativity $\mathcal{E}_f$ for $(a)$ $z=2$ fermionic Lifshitz tricritical point and $(b)$ Ising CFT at finite temperatures.}
At low temperatures $T \ll 1$, (a) $\mathcal{E}_f \sim \ln 2$ for the $z=2$ fermionic Lifshitz tricriticality (blue dotted line) and (b) $\mathcal{E}_f \sim -(c/2) \ln T$ for the Ising CFT (green dashed-dotted line) with the central charge $c=1/2$.
The red dashed lines guide the universal high-temperature power-law scaling of $\mathcal{E}_f (T \gg 1) \sim T^{-2}$ for a fermionic Gaussian state.} 
\label{fig: finiteTscaling} 
\end{figure}

At low temperatures, $T \ll 1$, the fermionic negativity for the Lifshitz tricritical point does not show noticeable temperature dependence and remains almost constant [Fig.~\ref{fig: finiteTscaling} (a)] while the other critical points described by CFTs exhibit the characteristic logarithmic temperature scaling, $\mathcal{E}_f \sim 2 \times (c/4) \ln \left[ \tanh (\pi l T)/ (\pi T) \right] \sim -(c/2) \ln T$ [Fig.~\ref{fig: finiteTscaling} (b)] \cite{Calabrese_finiteTnegativity, Shapourian_finiteTnegativity}.
The multiplicative factor of 2 comes from the two interfaces between the regions $A$ and $B$ when we impose periodic boundary conditions.
At sufficiently high temperatures, $T \gg 1$, not only the $z=2$ fermionic Lifshitz tricritical point but also for the other CFTs and the gapped phases show the universal power-law scaling behavior of the fermionic negativity $\mathcal{E}_f \sim T^{-2}$ (Fig.~\ref{fig: finiteTscaling}), which can be analytically shown via the high-temperature expansion of the reduced density operator (Appendix \ref{app: highT}).

Despite the partial transpose of the fermionic density operator being redefined as we discussed in Sec.~\ref{ssec: fermionic_partial_transpose}, the fermionic negativity $\mathcal{E}_f$ of the fermionic Lifshitz theory agrees with the bosonic negativity $\mathcal{E}_b$ of the Jordan-Wigner dual spin model with periodic boundary conditions \cite{oneway_QC, cluster_Ising, Ruben_1dSPT},
\begin{align}
H_{\mathrm{spin}} = \frac{1}{2}\sum_{n=1}^L X_n Z_{n+1} X_{n+2} - 2 X_n X_{n+1} - Z_n,
\label{eq: XZX-Ising}
\end{align}
where $X$ and $Z$ are the local Pauli operators.
The ground state at the tricritical point of the spin Hamiltonian $H_{\mathrm{spin}}$ is the $L$-body Greenberger-Horne-Zeilinger (GHZ) state, $|\Omega_{\mathrm{spin}}\rangle = \frac{1}{\sqrt{2}} \left[  \left | \uparrow \cdots \uparrow \right \rangle + \left |\downarrow \cdots \downarrow \right\rangle \right]$ \cite{crossing_qc}.
Hence, the bipartite negativity of the spin model is $\mathcal{E}_b = \ln 2$ and does not depend on the total length of the spin chain $L$ at $T=0$.
Since the low-lying domain wall excitations of the spin model can be effectively treated as a pair of noninteracting fermionic quasiparticles, the \emph{fermionic} $z=2$ Lifshitz theory yields a consistent prediction.

Notice that our findings differ from previous studies on bosonic $z=2$  Lifshitz theory in (1+1)-dimensions \cite{Lifshitz_boson, RK_separability}.
While the fermionic negativity $\mathcal{E}_f = \ln 2$ of the fermionic $z=2$ Lifshitz theory is constant, the bosonic negativity in the bosonic $z=2$ Lifshitz theory exhibits a logarithmic dependence on the system size, i.e., $\mathcal{E}_b \sim \ln L$ \cite{Lifshitz_boson}.
This disparity arises because fermionic and bosonic Lifshitz theories correspond to different actions.
To be more specific, our theoretical calculations are based on a specific lattice model realization of the $z=2$ fermionic Lifsthiz theory at $T=0$,
\begin{align}
S_f = \int_0^{\infty} \int_{-\infty}^{\infty}  \overline{\psi}\, \frac{\partial \psi}{\partial \tau} + \overline{\psi} \, \frac{\partial^2 \psi}{\partial x^2} \, dx \, d\tau.
\end{align}
This can be contrasted with the $z=2$ bosonic Lifshitz theory \cite{Lifshitz_boson, RK_separability},
\begin{align}
S_b = \frac{1}{2} \int_0^{\infty} \int_{-\infty}^{\infty} \left( \frac{\partial \varphi}{\partial \tau} \right)^2 + \varphi \frac{\partial^4 \varphi}{\partial x^4} \, dx \, d\tau.
\end{align}
Both actions, $S_b$ and $S_f$, are invariant under anisotropic dilation of space ($x \to \lambda x$) and time ($\tau \to \lambda^z \tau = \lambda^2 \tau$), with the appropriate field renormalization ($\varphi \to \sqrt{\lambda} \varphi$ and $\left(\psi, \overline{\psi}\right) \to \left(\psi, \overline{\psi}\right) / \sqrt{\lambda}$).
However, the precise forms of the actions are distinct, leading to different entanglement properties of the bosonic and fermionic Lifshitz theories.

\section{Entanglement negativity of spatially separated disjoint intervals \label{sec: disjoint}}

One of the most noticeable features of SPT phases is the presence of robust zero energy modes at the boundary.
The topologically protected edge modes manifest nontrivial bulk topology encoded in the ground state wavefunctions.
Therefore, it is natural to explore quantum entanglement between the two spatially separated blocks of fermions enclosing the edges \cite{Zeng_SSB_entropy, Zeng_SPT_entropy, topological_SE}.

\begin{figure}[h]
\usetikzlibrary {arrows.meta}
\begin{tikzpicture}[>=Stealth]
\draw[blue] (1.05, 0.8) node { $l$ };
\draw [<->, thick, blue] (-0.3, 0.5) -- (2.4, 0.5);
\draw[red] (6.65, 0.8) node { $l$ };
\draw [<->, thick, red] (5.3, 0.5) -- (8, 0.5);
\draw[] (3.85, 0.8) node { $d$ };
\draw [<->, thick] (2.4, 0.5) -- (5.3, 0.5);
\draw[blue] (1.05, -0.5) node { $A$ };
\draw[red] (6.65, -0.5) node { $B$ };
\draw[] (3.85, -0.5) node { $C$ };
\fill[semitransparent, blue] (-0.3,0.2) rectangle (2.4,-0.2);
\fill[semitransparent, red] (5.3,0.2) rectangle (8.0,-0.2);
\foreach \x in {0, 1.4, 2.8, 4.2, 5.6, 7} {
	\draw[draw=black, fill=white, thick] (\x, 0) circle (3pt);
	}
	
\foreach \y in {0.7, 2.1, 3.5, 4.9, 6.3, 7.7} {
	\filldraw (\y, 0) circle (3pt);
	}
\end{tikzpicture}
\caption{\textbf{Two colored disjoint intervals of $l$ unit cells separated by $d$ unit cells.}  The fermionic chain of length $L = 2l + d$ assumes open boundary conditions. Both intervals $A$ and $B$ include one of the edges.}
 \label{fig: disjoint_partition}
\end{figure}
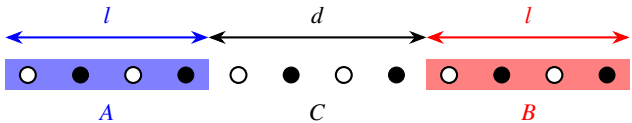

With a fixed system size $L$, we compute the fermionic negativity $\mathcal{E}_f$ between two intervals of length $l = (L-d)/2$ with a substantial separation $d$ (as illustrated in Fig.~\ref{fig: disjoint_partition}).
When $d\neq 0$, it becomes necessary to trace out the region $C$ between the regions $A$ and $B$.
This leads us to employ the mixed state entanglement measure for the reduced density operator $\rho = \mathrm{Tr}_C~|\Omega \rangle \langle \Omega |$, even if the ground state $|\Omega\rangle$ is prepared at $T=0$.
The fermionic negativity $\mathcal{E}_f$ is a suitable entanglement measure for quantifying the entanglement between these two disjoint blocks of fermions.

Considering the local nature of the Hamiltonian, it is reasonable to anticipate that the fermionic negativity tends towards zero with increasing separation $d$ between two blocks of fermions. 
This expectation has been rigorously verified for one-dimensional massive field theories \cite{negativity_massive} and CFTs for bosons \cite{Calabrese_QFTnegativity, Calabrese_negativity_extended}.
In the case of bosonic systems, such as the critical Ising chain, a critical separation $d_c$ exists such that the bosonic negativity drops to zero when the separation $d$ exceeds $d_c$ \cite{Ising_disjoint}.
In contrast, for fermionic CFTs, the fermionic negativity undergoes algebraic decay and eventually converges to zero as the separation approaches infinity \cite{fermion_CFT_disjoint}.

\begin{figure}[t]
  \centering
  \includegraphics{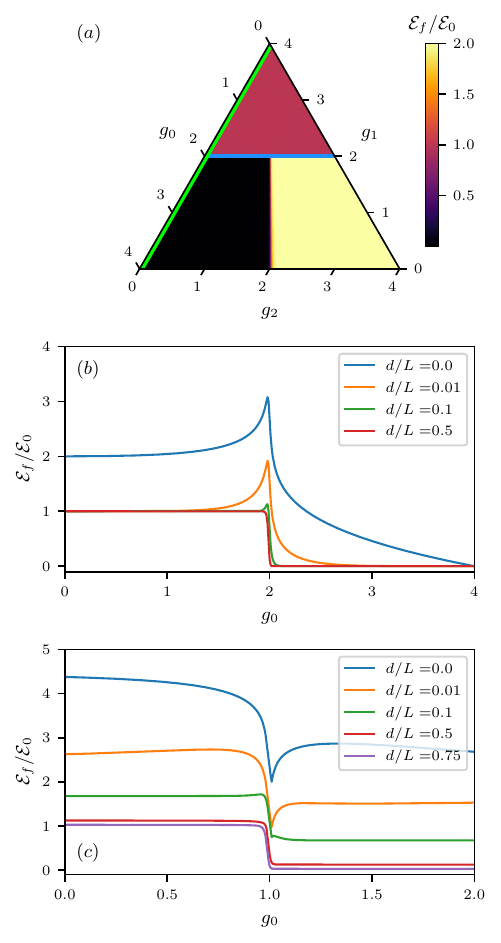}
  \caption{\textbf{Fermionic negativity $\mathcal{E}_f$ of the two disjoint intervals of length $l = (L-d)/2$ separated by the distance $d$ at zero temperature.} The open chain consists of $L = 200$ unit cells.
  (a) The negativity phase diagram ($d/L = 0.5$) reproduces the ground state phase diagram in Fig.~\ref{fig: phases}. The color scale denotes the normalized negativity $\mathcal{E}_f/\mathcal{E}_0$, which counts the number of Majorana zero modes per edge. $\mathcal{E}_0 = (\ln 2)/2$ is the fermionic negativity of two maximally entangled Majorana fermions.
  The fermionic negativities of the (b) Kitaev chain [green path in (a)] and (c) CFTs [blue path in (a)] show sharp steplike transition at the critical point and the quantized integer values elsewhere for the large separation $d$.}
  \label{fig: negativity_quantize}
\end{figure}

Nevertheless, for sufficiently large separations $d$, the fermionic negativity of topologically trivial fermionic CFTs becomes negligible as illustrated by the case where $g_0 > 1$ in Fig.~\ref{fig: negativity_quantize} (c). 
Hence, at $T=0$, the fermionic negativity of the disjoint intervals serves as an indicator of the number of Majorana zero modes.
This holds for both gapped and gapless topological states because the edge-to-edge entanglement remains finite irrespective of the separation $d$.
The pronounced, steplike increase in fermionic negativity at the phase boundary between the topological phase and the trivial phase [as shown in Fig.~\ref{fig: negativity_quantize} (b) and (c)] arises from the entanglement between topologically protected boundary modes.
However, at $T > 0$, the fermionic negativities for both gapped and gapless states vanish for large separation $d$ because the edge modes lose stability at finite temperatures, as previously discussed in Sec.~\ref{ssec: gapped_QCP}.

Earlier research has demonstrated that the conditional mutual information (or more generally, squashed entanglement \cite{squashed_entanglement}) between disjoint intervals can identify the gapped SPTs in spin models \cite{Zeng_SSB_entropy, Zeng_SPT_entropy} and fermionic systems \cite{topological_SE}.
Likewise, the fermionic negativity between the disjoint intervals differentiates not only the gapped fermionic SPTs but also the topologically distinct CFTs [Fig.~\ref{fig: negativity_quantize} (c)].
The negativity phase diagram [Fig.~\ref{fig: negativity_quantize} (a)] precisely reproduces the ground state phase diagram illustrated in Fig.~\ref{fig: phases}, all without the need to calculate any order parameter.

\section{Conclusion \label{sec: conclusion}}

In summary, we discussed how the fermionic negativity $\mathcal{E}_f$ can capture the finite temperature signatures of topological and quantum-critical ground states.
The fermionic negativity displays the quantum critical fan structure near the QCP between the gapped SPTs and the noticeable decrease at the fermionic Lifshitz tricritical point between the topologically distinct Ising CFTs.
The thermal states inside the quantum critical fan show logarithmic temperature dependence $\mathcal{E}_f \sim (c/2) \ln 1/T$ as CFTs at finite temperatures \cite{Calabrese_finiteTnegativity, Shapourian_finiteTnegativity}.
The low-temperature thermal states below the crossover temperature show nearly constant values due to the finite energy gap.
Although the $z=2$ fermionic Lifshitz tricritical point is gapless, its short-range correlation leads to the system size independent fermionic negativity $\mathcal{E}_f = \ln 2$ for the bipartite geometry with periodic boundary conditions.
Despite the zero spectral gap, thermal fluctuations do not significantly alter the constant entanglement at the Lifshitz tricritical point for moderate temperatures.
However, every thermal state of free fermions eventually exhibits universal $\mathcal{E}_f \sim T^{-2}$ temperature scaling in the high-temperature limit.

While our free fermion Hamiltonian exhibits Jordan-Wigner duality with the extended XZX cluster-Ising model \cite{Ruben_topology_fermion, Ruben_1dSPT, cluster_Ising, oneway_QC}, our analysis focuses solely on the 
entanglement negativity of fermionic Gaussian states.
Given that the refined definition of entanglement negativity has been introduced for fermions \cite{Shapourian_partial_TR}, it presents an intriguing avenue for future research to compare our results with the bosonic negativity of the Jordan-Wigner dual spin model.
However, due to the interacting nature of the spin model, conducting theoretical calculations becomes challenging.
Thus, the comparison between the two models is left for future work.
In addition, understanding how interactions among fermions modify the entanglement negativity of fermionic topological phases and QCPs is also an important question. 

\begin{acknowledgements}
We acknowledge support from the Technical University of Munich, the Deutsche Forschungsgemeinschaft (DFG, German Research Foundation) under Germany’s Excellence Strategy–EXC–2111–390814868, TRR 360 - 492547816, and from the European Research Council (ERC) under the European Unions Horizon 2020 research and innovation program (Grant Agreements No. 771537 and No. 851161), as well as the Munich Quantum Valley, which the Bavarian state government supports with funds from the Hightech Agenda Bayern Plus.
\end{acknowledgements}

\textit{Data and code availability:} Numerical data and simulation codes are available on Zenodo upon reasonable
request \cite{data}.

\appendix

\section{Covariance matrix of the BDI symmetry class \label{app: covariance}}

In Sec.~\ref{ssec: free_fermion_negativity}, we reviewed how to calculate the logarithmic negativity of a fermionic Gaussian state from the covariance matrix \cite{Peschel_RDMcorrelation, negativity_bound, Fagotti_XYdisjoint, Shapourian_finiteTnegativity, twisted_negativity}.
To calculate the covariance matrix, we need the block factorization of a real skew-symmetric matrix $A = Q\Sigma Q^T$ [Eq.~(\ref{eq: block_factorize_skewsymmetric})].
This appendix provides explicit formulae to evaluate the covariance matrix $\gamma$ by diagonalizing the Hamiltonians in the BDI symmetry class with a Majorana fermion basis.

Let us consider a quadratic Hamiltonian in the BDI symmetry class, which is a generalization of the $\alpha$-chain:
\begin{align}
H &=\frac{i}{2} \sum_{m,n=1}^{L} t_{mn} a_{m} b_{n},
\end{align}
where $a_m \equiv c_{2m -1}$ and $b_n \equiv c_{2n}$ are Majorana fermions as we defined for Eq.~(\ref{eq: alpha_chain}).
The Hamiltonian preserves both the time-reversal symmetry $\mathcal{T}$ and particle-hole symmetry $\mathcal{C}$, as discussed in Eqs.~(\ref{eq: C}) and (\ref{eq: T}).
There are only the terms between odd ($a_m \equiv c_{2m -1}$) and even ($b_n \equiv c_{2n}$) lattice sites because terms connecting two odd sublattice sites or two even sublattice sites such as $ia_m a_n$ or $ib_m b_n$ are forbidden by the chiral symmetry $\mathcal{S} = \mathcal{C}\mathcal{T}$.

The Hamiltonian can be diagonalized using the singular value decomposition of $t = U S V^T$:
\begin{align}
H &=\frac{i}{2} \sum_{m,n=1}^{L} \left( U S V^T \right)_{mn} a_{m} b_{n}
\nonumber \\
&= \frac{i}{2} \sum_{l=1}^{L} \varepsilon_l \left( \sum_{m=1}^{L} a_{m} U_{ml} \right) \left( \sum_{n=1}^{L} b_{n} V_{nl}\right) \equiv \frac{i}{2}\sum_{l=1}^{L} \varepsilon_l d_l' d_l''
\nonumber \\
&= \sum_{l=1}^{L} \varepsilon_l \left(f_l^\dagger f_l - \frac{1}{2}\right),
\label{eq: diagonalize_H}
\end{align}
where $\varepsilon_l \geq 0$ are the singular values of $t$, and $f_l^\dagger = \frac{1}{2} \left(d_l' - i d_l''\right)$ and $f_l = \frac{1}{2} \left(d_l' + i d_l''\right)$ are the creation and annihilation operators of the complex fermionic eigenmodes.

\begin{figure}[h]
  \centering
  \includegraphics{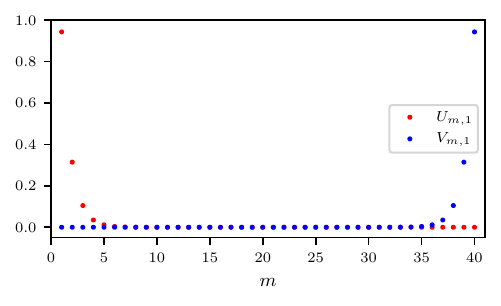}
  \caption{\textbf{Wavefunction amplitude of Majorana zero modes for the topological superconductor ($\omega = 1$) with $(g_0,g_1,g_2) = (1, 3, 0)$.} Open boundary conditions are assumed for a chain of $L = 40$ unit cells. Since $U_{m,1}$ and $V_{m,1}$ have sizable magnitudes only near the left ($1\leq m \leq 5$) and right ($35 \leq m \leq 40$) edges, $d_1' = \sum_m a_m U_{m,1}$ and $d_1'' = \sum_m b_m V_{m,1}$ are the two Majorana operators responsible for the topologically protected zero modes at the boundary.}
  \label{fig: edge_mode}
\end{figure}

In the case of SPT phases, the operators $d_1' = \sum_m a_m U_{m,0}$ and $d_1'' = \sum_m b_m V_{m,1}$ represent two Majorana zero-energy modes localized at the left and right edges, respectively.
As we can see from Fig.~\ref{fig: edge_mode}, $U_{m,1}$ and $V_{m,1}$ have noticeable magnitudes only near the left and right edges of the chain, respectively.
Notably, the creation of a complex fermion $f_1^\dagger = \frac{1}{2} \left(d_1' -id_1'' \right)$ incurs zero energy cost, specifically $\varepsilon_1 = 0$.
Hence, the degenerate ground state $f_1^\dagger|0\rangle$ is a linear combination of the Majorana zero modes localized at the left and right boundaries.
Consequently, the disjoint blocks of fermions that enclose these boundaries exhibit finite entanglement negativity, regardless of the degree of separation between them at $T=0$.

Using the singular values $\varepsilon_l$ and the left and right singular vectors $U$ and $V$, we can calculate the covariance matrix of Majorana fermions $\gamma_{jk} = \frac{i}{2} \left \langle [ c_j , c_k ] \right\rangle = \left \langle i c_j c_k \right \rangle - i \delta_{jk}$.
Due to the chiral symmetry $\mathcal{S} = \mathcal{C}\mathcal{T}$, $\gamma_{2m-1,2n-1} = \gamma_{2m, 2n} = 0$ for all $1 \leq m,n \leq L$.
The nonvanishing matrix elements are
\begin{align}
\gamma_{2m-1, 2n} &= -\gamma_{2n, 2m-1} = \sum_{k=1}^{L} U_{mk} V_{nk} \left( 2 \left \langle f_k^\dagger f_k \right\rangle - 1 \right)
\\
&= - \sum_{k=1}^{L} U_{mk} V_{nk} \tanh \left( \frac{\varepsilon_k}{2T} \right) \xrightarrow{T\to0} - \left(UV^T \right)_{mn},
\end{align}
where $\left \langle f_k^\dagger f_k \right\rangle = \left( e^{\varepsilon_k/T} +1 \right)^{-1}$ obeys the Fermi-Dirac statistics at finite temperatures $T > 0$.

\section{Trace formulae for a fermionic Gaussian state \label{app: trace}}

This appendix shows how to calculate $\mathrm{Tr}\, \rho^\alpha$ for any $\alpha \in \mathbb{R}$ using the covariance matrix $\gamma$ \cite{Shapourian_partial_TR, negativity_bound}.
The formulae below are especially useful for a mixed state because we can avoid calculations for the entanglement Hamiltonian $\mathcal{H}$ of the reduced density operator $\rho= \mathrm{Tr}_{B}\, \rho_{A\cup B} = e^{-\mathcal{H}} / Z$ after the partial trace $\mathrm{Tr}_B$.
We only need to compute the covariance matrix $\gamma$ for the region of interest from the physical Hamiltonian $H$.

Let us consider a Gaussian state of Majorana fermions $c_j$
\begin{align}
\rho &= \frac{1}{Z} e^{-\frac{i}{4} \sum_{j,k=1}^{2N} A_{jk} c_j c_k}
= \frac{1}{Z} e^{-\sum_{n=1}^{N} \lambda_n \left( f_n^\dagger f_n - \frac{1}{2} \right)}
\\
&=\prod_{n=1}^N \frac{e^{-\lambda_n f_n^\dagger f_n}}{1 + e^{-\lambda_n}}
=\prod_{n=1}^N \left[1 - \mathcal{F}\left(\lambda_n \right)\right] e^{-\lambda_n f_n^\dagger f_n}
\end{align}
where $A_{jk} = -A_{kj} \in \mathbb{R}$, $\lambda_n$ are eigenvalues of $iA$, and $\mathcal{F}(x) = 1/\left(e^x + 1 \right)$ is the Fermi-Dirac function.
Due to Eq.~(\ref{eq: expiA}), $e^{-\lambda_n}$ is equal to the eigenvalue of $\left(\mathbb{I} - i\gamma\right)\left(\mathbb{I} + i\gamma\right)^{-1}$.
Starting from Eq.~(\ref{eq: gamma_factorize}), we can simultaneously diagonalize $\mathbb{I} \pm i\gamma$ with a unitary matrix $U$ and an orthogonal matrix $Q$:
\begin{align}
\frac{\mathbb{I}+i\gamma}{2} &=Q U \left[
\bigoplus_{n=1}^{N}
\begin{pmatrix}
1 - \mathcal{F}(\lambda_n) & 0
\\
0 & \mathcal{F}(\lambda_n)
\end{pmatrix}
\right]U^\dagger Q^T
\\
&\equiv Q U \left[
\bigoplus_{n=1}^{N}
\begin{pmatrix}
1 - \zeta_n & 0
\\
0 & \zeta_n
\end{pmatrix}
\right]U^\dagger Q^T,
\\
\frac{\mathbb{I}-i\gamma}{2} &=Q U \left[
\bigoplus_{n=1}^{N}
\begin{pmatrix}
\mathcal{F}(\lambda_n) & 0
\\
0 & 1-\mathcal{F}(\lambda_n)
\end{pmatrix}
\right]U^\dagger Q^T,
\end{align}
where $\zeta_n = \mathcal{F}\left(\lambda_n\right) =  1 / \left( e^{\lambda_n} + 1 \right)$ are eigenvalues of the Hermitian matrix $\left(\mathbb{I}+i\gamma\right)/2$.

Then, we can calculate $\mathrm{Tr}\,\rho^\alpha$ from the eigenvalues of $\left(\mathbb{I}+i\gamma\right)/2$:
\begin{align}
\rho^\alpha &= \prod_{n=1}^N \left[1 - \mathcal{F}\left(\lambda_n\right) \right]^\alpha e^{-\alpha \lambda_n f_n^\dagger f_n}
\\
\Rightarrow \mathrm{Tr}\, \rho^\alpha \,&= \prod_{n=1}^N \left[ 1- \mathcal{F}\left(\lambda_n\right)\right]^\alpha \left(1+e^{-\alpha\lambda_n}\right)
\nonumber \\
&= \prod_{n=1}^N \left[1-\mathcal{F}\left(\lambda_n\right)\right]^\alpha + \left[\mathcal{F}\left(\lambda_n\right)\right]^\alpha
\\
&= \prod_{n=1}^N \left(1-\zeta_n \right)^\alpha + \zeta_n^\alpha.
\end{align}
Since $0 \leq \mathcal{F}(\lambda_n), 1 - \mathcal{F}(\lambda_n) \leq 1$ for any $\lambda_n \geq 0$,
\begin{align}
\mathrm{Tr} \, \rho^\alpha = \sqrt{\det \left[\left(\frac{\mathbb{I}-i\gamma}{2} \right)^\alpha +\left(\frac{\mathbb{I}+i\gamma}{2} \right)^\alpha \right]}.
\end{align}
Hence,
\begin{align}
\mathrm{Tr}\, \rho^2 &= \sqrt{\det \left[\left(\frac{\mathbb{I}-i\gamma}{2} \right)^2 +\left(\frac{\mathbb{I}+i\gamma}{2} \right)^2 \right]}
\nonumber \\
&= \sqrt{\det \left( \frac{\mathbb{I} - \gamma^2}{2}\right)}
=\sqrt{\det \left(\frac{\mathbb{I} - \gamma_{+} \gamma_{-}}{2} \right)} =\frac{Z_{\times}}{Z_{+}Z_{-}},
\end{align}
where $Z_\sigma = \mathrm{Tr} ~e^{-\frac{i}{4}\sum_{jk} W_{jk}^{\sigma} c_j c_k} = \mathrm{Pf}  \left(\mathbb{I} + e^{-iW^{\sigma}} \right)$ for $\sigma = +, -, \times$, and the skew-symmetric matrices $W^{\sigma}$ are defined in Sec.~\ref{ssec: product_covariance}.
Brute force calculations can check the third equality.
The last equality holds because
\begin{align}
\left(\frac{Z_{\times}}{Z_{+}Z_{-}}\right)^2 &=
\frac{\det \left( \mathbb{I} + e^{-iW^{\times}}\right)}{\det \left( \mathbb{I} + e^{-iW^{+}}\right) \det \left(\mathbb{I} + e^{-iW^{-}} \right)}
\\
&=\frac{\det\left( \frac{\mathbb{I} + i\gamma_{+}}{2}\right)\det\left( \frac{\mathbb{I} + i\gamma_{-}}{2}\right)}{\det\left( \frac{\mathbb{I} + i\gamma_{-}}{2}\right)\det \left[ \left(\frac{\mathbb{I} -\gamma_{+}\gamma_{-}}{2}\right)^{-1}\right]\det\left( \frac{\mathbb{I} + i\gamma_{+}}{2}\right)}
\label{eq: ZxZ+Z-derivation}
\\
&=\det\left(\frac{\mathbb{I} - \gamma_{+}\gamma_{-}}{2}\right).
\end{align}
To get Eq.~(\ref{eq: ZxZ+Z-derivation}), we used Eq.~(\ref{eq: gamma_x}) and the identity
\begin{align}
\mathbb{I} + e^{-iW^{\sigma}} = \mathbb{I} + \frac{\mathbb{I}-i\gamma_{\sigma}}{\mathbb{I}+i\gamma_{\sigma}}
= \left(\frac{\mathbb{I} + i\gamma_{\sigma}}{2}\right)^{-1}.
\end{align}

\section{Fermionic negativity of the maximally entangled two Majorana zero modes \label{app: E0}}

Suppose we have two Majorana zero modes $d_0'$ and $d_0''$ localized at the left and right ends of the chain, respectively.
Since $f^\dagger = \frac{1}{2} \left( d_0' - id_0'' \right)$ can create the zero energy edge mode, the ground state of the open fermionic chain is two-fold degenerate with two different fermion parity $\left \langle i d_0' d_0'' \right \rangle = \left \langle 2 f^\dagger f - 1 \right \rangle  = \pm 1$.
As $\left ( i d_0' d_0'' \right)^2 = 1$, the ground state saturates the maximum correlation between the  Majorana zero modes.
In this sense, the two Majorana zero modes are maximally entangled.

Let us focus on the case $\left \langle i d_0' d_0'' \right \rangle = +1$.
Then the covariance matrix for the two-Majorana-fermion system is
\begin{align}
\gamma =
\begin{pmatrix}
0 & 1
\\
-1 & 0
\end{pmatrix}.
\end{align}
If we take the fermionic partial transpose with respect to the Majorana zero modes localized at the left end, Eqs.~(\ref{eq: gamma_pm}) and (\ref{eq: gamma_x}) give
\begin{align}
\gamma_{\pm} =
\begin{pmatrix}
0 & \pm i
\\
\mp i & 0
\end{pmatrix}
\Rightarrow
\frac{1}{2} \left( \mathbb{I} + i\gamma_{\times}\right) = \frac{\mathbb{I}}{2}.
\end{align}
Then, by Eq.~(\ref{eq: negativity_formula_eigenvalue}),
\begin{align}
\mathcal{E}_0 = \ln \left( \sqrt{1 - \frac{1}{2}} + \sqrt{\frac{1}{2}} \right) = \ln \sqrt{2} = \frac{1}{2}\ln 2.
\end{align}
Note that $\mathcal{E}_0$ is the logarithm of the quantum dimension of Majorana fermion $\mathcal{D} = \sqrt{2}$.

If there are $n$ Majorana zero modes per edge, the total fermionic negativity is just $n\mathcal{E}_0$ because the fermionic negativity respects strong additivity for different flavors of fermions [Eq.~(\ref{eq: additive})].

%

%

\section{Fermionic $z=2$ Lifshitz tricritical point at $T=0$ \label{app: Lifshitz}}

This appendix discusses our new analytical result about the fermionic negativity of the $z=2$ fermionic Lifshitz tricritical point at zero temperature from the covariance matrix.
With periodic boundary conditions, we define the discrete Fourier transformations
\begin{align}
a_n &= \sqrt{\frac{2}{L}}\sum_{k = -\frac{L}{2}}^{\frac{L}{2}-1} e^{2\pi i kn/L} a_k
\equiv \sqrt{\frac{2}{L}} \sum_q e^{iqn} a_q,
\\
a_q & = \frac{1}{\sqrt{2L}} \sum_{n=1}^{L} e^{- i qn} a_n
\end{align}
with $q = 2\pi k /L$.
Then, the critical Hamiltonian $H$ can be diagonalized as follows:
\begin{align}
H &= \frac{i}{2} \sum_{n=1}^L a_n b_n - 2 a_{n+1} b_n + a_{n+2} b_n
\\
&=
\frac{1}{2}\sum_q
\begin{pmatrix}
a_q^\dagger, & b_q^\dagger
\end{pmatrix}
U_q
\begin{pmatrix}
\epsilon_q & 0
\\
0 & -\epsilon_q
\end{pmatrix}
U_q^\dagger
\begin{pmatrix}
a_q 
\\
b_q
\end{pmatrix},
\end{align}
where $\varepsilon_q = 2(1- \cos q)$, and
\begin{align}
U_q = \frac{1}{\sqrt{2}}
\begin{pmatrix}
1 & 1 \\
ie^{iq} & -ie^{iq}
\end{pmatrix}.
\end{align}
For small $q \ll 1$, we can check $\varepsilon_q \approx q^2$.

Using the eigenvectors $U_q$ of $H$, we can calculate the nonvanishing covariance matrix elements $\gamma_{2m - 1, 2n} = -\gamma_{2n, 2m - 1} = \langle i a_m b_n \rangle = \delta_{m,n+1}$ for $1\leq m, n \leq L$, and $L+1 \equiv 1$ for periodic boundary conditions.
Notice that the correlation between Majorana fermions is extremely short-ranged even though the bulk is gapless.
This short-range correlation is the physical origin of the nondiverging, constant entanglement negativity of the fermionic Lifshitz theory.

The explicit expression for the covariance matrix $\gamma$ is
\begin{align}
\gamma &=
\begin{pmatrix}
0 &  &  &  &  &  &  &  & 1
\\
& 0 & -1 
\\ 
& 1 & 0 
\\ 
&  &  & 0 & -1  
\\ 
&  &  & 1 & 0 
\\ 
&  &  &  &  & \ddots
\\ 
&  &  &  &  &  & 0 & -1
\\ 
&  &  &  &  &  &  1 & 0
\\ 
-1 &  &  &  &  &  &  &  & 0
\end{pmatrix}
\\
&\equiv
\begin{pmatrix}
0 & \cdots & 0 & 0 & \cdots & 1
\\
\vdots & -iY & \vdots & \vdots & &\vdots
\\
0 & \cdots & 0 & -1 & \cdots & 0
\\
0 & \cdots & 1 & 0 & \cdots & 0
\\
\vdots & & \vdots & \vdots & -iY & \vdots
\\
-1 & \cdots & 0 & 0 & \cdots & 0
\end{pmatrix},
\label{eq: gamma_Lifshitz}
\end{align}
where $-iY \equiv \bigoplus_{n=1}^{L/2 - 1}
\begin{pmatrix}
0 & -1
\\ 1 & 0
\end{pmatrix}$, and the vertical and horizontal dots denote strings of zeros.

\subsection{Bipartite geometry \label{ssec: Lifshitz_bipartite}}
Let us consider an equal bipartition of the fermionic chain (Fig.~\ref{fig: bipartition}).
If we are interested in the entanglement between the two equal partitions at $T=0$, the second R\'{e}nyi entropy $S_2 = 0$. 
Hence, we only need to calculate the eigenvalues of $\frac{1}{2}\left( 1 + i\gamma_\times\right)$ in Eq.~(\ref{eq: gamma_x}). 
From Eq.~(\ref{eq: gamma_Lifshitz}), we can write the covariance matrices $\gamma_{\pm}$ for the fermionic partial transpose of the density operator $\rho^{R_A}$ and its Hermitian conjugate $(\rho^{R_A})^\dagger$:
\begin{align}
\gamma_{\pm} &=
\begin{pmatrix}
0 & \cdots & 0 & 0 & \cdots & \pm i
\\
\vdots & iY & \vdots & \vdots & &\vdots
\\
0 & \cdots & 0 & \mp i & \cdots & 0
\\
0 & \cdots & \pm i & 0 & \cdots & 0
\\
\vdots & & \vdots & \vdots & -iY & \vdots
\\
\mp i & \cdots & 0 & 0 & \cdots & 0
\end{pmatrix}.
\end{align}
Note that the brute force matrix multiplications give $\mathbb{I} - \gamma_{+} \gamma_{-} = 2\mathbb{I}$, which significantly simplifies the further calculations.
Then,
\begin{align}
\frac{\mathbb{I} + i\gamma_{\times}}{2}
&= \frac{\mathbb{I} + i\gamma_{-}}{2} \left( \frac{\mathbb{I} - \gamma_{+} \gamma_{-}}{2} \right)^{-1} \frac{\mathbb{I} + i\gamma_{+}}{2}
\\
&= \frac{1}{4} \left( \mathbb{I} - \gamma_{-} \gamma_{+} + i\gamma_{+} +i\gamma_{-} \right)
\\
&=\frac{1}{2}
\begin{pmatrix}
1 & \cdots  & 0 & 0 & \cdots & 0
\\ \vdots & \mathbb{I} - Y  & \vdots & \vdots & & \vdots
\\ 0 & \cdots & 1 & 0 & \cdots & 0
\\ 0 & \cdots & 0 & 1 & \cdots & 0
\\ \vdots & & \vdots & \vdots & \mathbb{I} + Y & \vdots
\\ 0 & \cdots & 0 & 0 & \cdots & 1
\end{pmatrix}.
\end{align}

Since the eigenvalues of $\frac{1}{2}\left ( \mathbb{I} \pm Y\right)$ are $0$ and $1$,  the spectrum of $\frac{1}{2}\left(\mathbb{I}+i\gamma_{\times}\right)$ is $\left \{ \frac{1}{2}, \frac{1}{2}, \frac{1}{2}, \frac{1}{2}, 0, \cdots, 0, 1, \cdots, 1 \right \}$, where the eigenvalue $\frac{1}{2}$ is four-fold degenerate, and $0$ and $1$ are $(L - 2)$-fold degenerate.
From Eq.~(\ref{eq: negativity_formula_eigenvalue}), the fermionic negativity of the Lifshitz tricritical point is
\begin{align}
\mathcal{E}_f &= \frac{1}{2} \Bigg[
4 \ln\left(\sqrt{1 - \frac{1}{2}} + \sqrt{\frac{1}{2}} \, \right)
\nonumber \\
&+ (L-2)\left\{ \ln\left(\sqrt{1 - 0} + \sqrt{0}\,  \right)
+ \ln\left(\sqrt{1 - 1} + \sqrt{1} \, \right) \right\}
\Bigg]
\nonumber \\
&= \ln 2.
\end{align}
Since the correlation between Majorana fermions is short-ranged, the intervals $A$ and $B$ are entangled only at the interface.
Thus, the fermionic negativity $\mathcal{E}_f$ of the Lifshitz tricritical point is independent of the system size $L$.

\subsection{Disjoint intervals}
\begin{figure}[h]
\usetikzlibrary {arrows.meta}
\begin{tikzpicture}[>=Stealth]
\draw[blue] (1.8, 0.8) node { $l$ };
\draw [<->, thick, blue] (1.15, 0.5) -- (2.35, 0.5);
\draw[red] (5.9, 0.8) node { $l$ };
\draw [<->, thick, red] (5.35, 0.5) -- (6.55, 0.5);
\draw[blue] (1.8, -0.5) node { $A$ };
\draw[red] (5.9, -0.5) node { $B$ };
\draw[black] (3.85, -0.5) node { $C$ };
\draw[black] (0.45, -0.5) node { $D$ };
\draw[black] (7.25, -0.5) node { $D$ };
\draw[black] (3.85, 0.8) node { $d$ };
\draw [<->, thick, black] (2.35, 0.5) -- (5.35, 0.5);
\fill[semitransparent, blue] (1.15,0.2) rectangle (2.35,-0.2);
\fill[semitransparent, red] (5.35,0.2) rectangle (6.55,-0.2);
\foreach \x in {0, 1.4, 2.8, 4.2, 5.6, 7} {
	\draw[draw=black, fill=white, thick] (\x, 0) circle (3pt);
	}
	
\foreach \y in {0.7, 2.1, 3.5, 4.9, 6.3, 7.7} {
	\filldraw (\y, 0) circle (3pt);
	}
\end{tikzpicture}
\caption{\textbf{Two disjoint intervals $A$ and $B$ of length $l$ are separated by the distance $d$.}
The total length of the fermionic chain $L$ is assumed to be much greater than $2l + d$.}
 \label{fig: disjoint_tripartite}
\end{figure}

We examine the fermionic negativity of the two disjoint intervals $A$ and $B$ (Fig.~\ref{fig: disjoint_tripartite}).
To ensure the intervals are spatially separated with periodic boundary conditions, the intervals $C$ and $D$ must be non-empty.
Then, the reduced covariance matrix $\tilde{\gamma}$ for the mixed state $\rho = \mathrm{Tr}_{C\cup D} \, |\Omega \rangle \langle \Omega |$ can be written as
\begin{align}
\tilde{\gamma} &=
\begin{pmatrix}
0 & \cdots & 0 & 0 & \cdots & 0
\\
\vdots & -i\bar{Y} & \vdots & \vdots & &\vdots
\\
0 & \cdots & 0 & 0 & \cdots & 0
\\
0 & \cdots & 0 & 0 & \cdots & 0
\\
\vdots & & \vdots & \vdots & -i\bar{Y} & \vdots
\\
0 & \cdots & 0 & 0 & \cdots & 0
\end{pmatrix}.
\end{align}

By calculating the eigenvalues of $(\mathbb{I} + i\tilde{\gamma})/2$ and $(\mathbb{I} + i\tilde{\gamma}_{\times})/2$, we get $S_2 = 2\mathcal{E}_{\times} = 2\ln 2$ at $T = 0$.
The numerical calculations also confirmed that $S_2 = 2 \mathcal{E}_{\times}$ for any $T > 0$.
Therefore, the fermionic negativity for the disjoint intervals is always $\mathcal{E}_f = -\frac{1}{2}S_2 + \mathcal{E}_{\times} = 0$ for any $d > 0$.
This results from ultra-short-ranged correlations of the $z=2$ fermionic Lifshitz theory in one spatial dimension.

\section{Entanglement negativity of free fermions at high temperatures \label{app: highT}}

At sufficiently high temperatures $T \gg 1$, the numerical calculations show the universal temperature dependence of the fermionic negativity, $\mathcal{E}_f \sim T^{-2}$, for all quantum states including gapped SPTs, CFTs, and the Lifshitz critical point.
This appendix explains the universal scaling of the fermionic Gaussian state using the high-temperature expansion of the density operator:
\begin{align}
\rho = \frac{1}{Z} e^{-\beta H} = \frac{1}{Z} \left( 1 - \beta H + \frac{\beta^2}{2} H^2 + \cdots \right).
\end{align}
To simplify the notation, let us denote $\widetilde{H} \equiv H^{R_A}$.
Likewise, we denote $\widetilde{H^2} \equiv \left(H^2\right)^{R_A}$.
Then,
\begin{widetext}
\begin{align}
&\rho^{R_A} \left( \rho^{R_A}\right)^\dagger = 
\frac{1}{Z^2} \left[ \mathbb{I} -\beta \left( \widetilde{H}+ \widetilde{H}^\dagger \right) + \frac{\beta^2}{2} \left( \widetilde{H^2} + \widetilde{H^2}^\dagger + 2 \widetilde{H} \widetilde{H}^\dagger \right) + \mathcal{O}(\beta^3)\right]
\\
\Rightarrow
&\sqrt{\rho^{R_A} \left( \rho^{R_A}\right)^\dagger }
= \frac{1}{Z}
\left[
\mathbb{I} - \frac{1}{2} \left (
\beta \left(\widetilde{H} + \widetilde{H}^\dagger  \right)
-\frac{\beta^2}{2} 
\left( \widetilde{H^2} + \widetilde{H^2}^\dagger + 2 \widetilde{H} \widetilde{H}^\dagger \right)
\right )
-\frac{\beta^2}{8} \left(
\widetilde{H}^2 + \left(\widetilde{H}^\dagger\right)^2 + \widetilde{H}\widetilde{H}^\dagger + \widetilde{H}^\dagger\widetilde{H}
\right) + \mathcal{O}(\beta^3)
\right]
\nonumber
\\
\Rightarrow 
&\, \mathrm{Tr}\, \sqrt{\rho^{R_A} \left( \rho^{R_A}\right)^\dagger } 
= \frac{d}{Z}
\left[
1
-\frac{\beta}{2d}
\mathrm{Tr}\, 
\left ( \widetilde{H} + \widetilde{H}^\dagger \right)
+ \frac{\beta^2}{4} \mathcal{C} + \mathcal{O}(\beta^3)
\right],
\label{eq: highTexpansion}
\end{align}
\end{widetext}
where $d = \mathrm{Tr}\,\mathbb{I} = 2^L$ for a chain of $2L$ Majorana fermions, and
\begin{multline}
\mathcal{C} = \frac{1}{2d} \mathrm{Tr} \left[
2\left(\widetilde{H^2} + \widetilde{H^2}^\dagger + 2\widetilde{H} \widetilde{H}^\dagger 
\right)\right.
\\
\left.-\left(
\widetilde{H}^2 + \left(\widetilde{H}^\dagger\right)^2 + \widetilde{H}\widetilde{H}^\dagger + \widetilde{H}^\dagger\widetilde{H}\right)
\right].
\end{multline}

Given a general Hamiltonian of noninteracting Majorana fermions $H = \frac{i}{4} \sum_{jk} h_{jk} c_j c_k$,
its fermionic partial transpose
\begin{multline}
H^{R_A} = \widetilde{H} =  -\frac{i}{4}\sum_{j,k\in A} h_{jk} c_j c_k + \frac{i}{4} \sum_{j,k \in B} h_{jk} c_j c_k
\\
- \frac{1}{4} \sum_{j \in A, k \in B} h_{jk} c_j c_k
- \frac{1}{4} \sum_{j \in B, k \in A} h_{jk} c_j c_k
\end{multline}
by the definition in Eq.~(\ref{eq: negativity_def}).
Since $\mathrm{Tr} \, \left( i c_j c_k \right) = 0$,
\begin{multline}
 \mathrm{Tr}\, \left( \widetilde{H} + \widetilde{H}^\dagger \right)
 =  \frac{1}{2} \mathrm{Tr}\, \left[
 -\sum_{j,k \in A} h_{jk} ic_j c_k + \sum_{j,k \in B} h_{jk} ic_j c_k
 \right] = 0.
 \nonumber
\end{multline}
From Eq.~(\ref{eq: highTexpansion}),
\begin{align}
\mathcal{E}_f &= \ln \mathrm{Tr}\, \sqrt{\rho^{R_A} \left( \rho^{R_A} \right)^\dagger}
\approx \ln \left(1 + \beta^2 \mathcal{C} \right) + \ln \left(d/Z\right)
\nonumber \\
&\approx \beta^2 \mathcal{C} + L \ln 2 - \ln Z.
\label{eq: high_T_Ef_derivation}
\end{align}

For noninteracting fermions, we can calculate the partition function $Z$ exactly:
\begin{align*}
Z &= \mathrm{Tr}\, e^{-\beta H} 
= \sum_{\{n_l\}} \left \langle \{n_l\} \right | e^{-\beta\sum_l \left(f_l^\dagger f_l - \frac{1}{2} \right)} \left | \{n_l\}  \right \rangle
\\
&= \prod_{l=1}^L  \sum_{n_l = 0, 1}e^{-\beta\varepsilon_l \left( n_l - \frac{1}{2}\right)} = \prod_{l=1}^L 2 \cosh \left( \frac{\beta\varepsilon_l}{2}\right).
\end{align*}
Hence,
\begin{align*}
\ln Z &= L \ln 2 + \sum_{l=1}^L \ln \cosh \left( \frac{\beta \varepsilon_l}{2}\right)
\\
&= L \ln 2 + \sum_{l=1}^{L} \ln \left( 1 + \frac{\beta^2 \varepsilon_l^2}{4} + \cdots \right)
\\
&\approx L \ln 2 + \beta^2 \sum_{l=1}^L \frac{\varepsilon_l^2}{4} + \mathcal{O}(\beta^4).
\end{align*}
From Eq.~(\ref{eq: high_T_Ef_derivation}), we get
\begin{align}
\mathcal{E}_f = \frac{\beta^2}{4} \left( \mathcal{C} - \sum_{l=1}^L \varepsilon_l^2 \right) + \mathcal{O}(\beta^3).
\end{align}
Thus, the bipartite fermionic negativity shows the universal high-temperature scaling $\mathcal{E}_f \propto T^{-2}$ for any fermionic Gaussian state.

\bibliography{negativity}

\end{document}